\documentstyle[emulateapj]{article}

\lefthead{Alonso-Herrero et al.}
\righthead{The non-stellar infrared continuum of Seyfert galaxies}

\begin{document}
\title{The non-stellar infrared continuum of Seyfert galaxies
\footnote{This work is based on observations collected at UKIRT. The UKIRT
is operated by the Joint Astronomy Centre on behalf of the 
UK Particle Physics and Astronomy Research Council. Based
on observations with the NASA/ESA Hubble
Space Telescope, obtained at the Space Telescope Science Institute, which
is operated by the Association of Universities for Research in Astronomy,
Inc. under NASA contract No. NAS5-26555.} 
}

\author{Almudena Alonso-Herrero}

\affil{Steward Observatory, The University of Arizona, Tucson, AZ 85721, USA}
\affil{Present address: University of Hertfordshire, Department 
of Physical Sciences, College Lane, Hatfield, Herts AL10 9AB, UK}

\author{Alice C. Quillen}

\affil{Steward Observatory, The University of Arizona, Tucson, AZ 85721, USA}

\author{Chris Simpson}

\affil{Subaru Telescope, National Astronomical Observatory of Japan, 
650 N. A`Oh\={o}k\={u} Place, Hilo, HI 96720, USA}

\author{Andreas Efstathiou}

\affil{Astrophysics Group, Imperial College London, Blackett Laboratory, 
Prince Consort Rd, London SW7 2BZ, UK}

\and

\author{Martin J. Ward}

\affil{Department of Physics and Astronomy, University of Leicester, 
Leicester LE1 7RH, UK}

\begin{abstract}
$JHKL^\prime M$ ($1-5\,\mu$m) 
imaging of a sample of Seyfert 2 galaxies is presented. 
We have performed an accurate estimate of the
near-infrared non-stellar nuclear fluxes. We confirm that the 
near-infrared nuclear continuum between 1 and $2.2\,\mu$m of some 
Seyfert 2s is dominated by stellar emission, whereas the continuum emission 
at longer wavelengths ($\lambda = 3-5\,\mu$m) is 
almost entirely non-stellar in origin.  The non-stellar spectral
energy distributions (SED) in the infrared (up to $15\,\mu$m)  of 
Seyfert galaxies  show
a variety of shapes, and they are well reproduced with 
the tapered disk models of Efstathiou \& Rowan-Robinson (1995). 
We have used two models, one including an optically thin cone component 
found to fit the SED of NGC~1068, and a coneless model. Although our modelling
of the SEDs does not allow us to favor either model to account for 
all the observed SEDs, we find that 
the viewing angle towards the central source is well constrained by
both models. The galaxies in our sample have fitted values of 
the viewing angle in the range $\theta_{\rm v} = 0\arcdeg-64\arcdeg$, 
for the assumed model parameters. 
We have also investigated non-stellar color-color diagrams 
($L^\prime-M$ vs. $H-M$ and $L^\prime-M$ vs. $H-L^\prime$). The colors of 
the Seyfert galaxies with viewing angles $\theta_{\rm v} 
< 30\arcdeg$ are better reproduced with the cone model. These
diagrams  provide a good means to separate Seyfert 2s with 
moderate obscuration ($A_V \lesssim 20\,$mag from 
hard X-ray observations) from those with high obscuration.

The ground-based $4.8\,\mu$m and
ISO $9.6\,\mu$m luminosities  are well correlated with the 
hard X-ray luminosities of Seyfert 1s and 2s. These 
continuum emissions appear as a good indicator of the AGN 
luminosity, at least in the cases of hard X-ray Compton-thin 
Seyfert galaxies ($N_{\rm H} \le 10^{24}\,{\rm cm}^{-2}$).
We finally stress the finding that some Compton thick 
galaxies show bright non-stellar emission at $5\,\mu$m. This suggests 
that the near-infrared emission in Seyfert galaxies is produced in 
an extended component illuminated by the central source, that is more visible
from all viewing angles, providing a good explanation for the  
differing $N_{\rm H}$/$A_V$ ratios found in some Seyfert 2s. 
We discuss possible implications of mid-infrared 
surveys for the search of counterparts of
highly obscured hard X-ray sources.

\end{abstract}


\section{INTRODUCTION}
There is now considerable evidence that 
the infrared emission in active galaxies (Seyfert galaxies and 
radio quiet quasars) 
is thermal in origin, namely hot dust emission
(e.g., Rieke \& Low 1975; Rieke 1978; Barvainis 1987;
McAlary \& Rieke 1988; Sanders 
et al. 1989; Danese et al. 1992 and 
recently Andreani, Franceschini, \& Granato 1999; 
Haas et al. 2000; Polletta et al. 2000  and 
references therein). The strong mid and far infrared emission in active
galactic nuclei is often interpreted as produced by a disk-like 
(Sanders et al. 1989) or torus-like (e.g., Pier \& Krolik 1992; Granato 
\& Danese 1994; Efstathiou \& Rowan-Robinson 1995 among others)
dust configuration heated by the central source. This obscuring 
material which 
hides a direct view of the central source in type 2 objects, absorbs  
a large fraction of the X-ray/UV/optical 
energy emitted by the central source and reradiates it in the 
infrared spectral range. 

The nature of the non-stellar infrared continuum in Seyfert 2 galaxies  
where the stellar component dominates the nuclear near-infrared emission 
(Edelson \& Malkan 1986; Alonso-Herrero, 
Ward, \& Kotilainen 1996), has eluded us due to the difficulty in estimating
the  non-stellar contribution at near-infrared wavelengths (up to 
$2\,\mu$m). 
High resolution near-infrared images observed with 
NICMOS on the {\it HST} have now been used to
detect and measure the nuclear non-stellar emission in Seyfert galaxies
(Quillen et al. 2000b). This work reported 
that almost 100\% of all Seyfert 1--1.9 galaxies and 
50\% of all Seyfert 2s show point sources at $1.6\,\mu$m. This
unresolved emission is variable in some of
the Seyfert 2s and thus is presumably non-stellar emission associated
with the central engine (Quillen et al. 2000a). 
The high detection rate of point sources in the near-infrared 
is in contrast with the small 
number of point sources reported in Seyfert 2s in the {\it HST} optical 
survey of Malkan, Gorjian, \& Tam (1998).
Previous attempts to
model  the non-stellar spectral energy distributions (SEDs) of Seyfert 
2s (e.g., Edelson \& Malkan 1986, and recently Fadda et al. 1998, and 
references therein) suffered from a number of problems, including 
large aperture photometry, and uncertain
determinations of the stellar component which may dominate the near-infrared
emission up to $2.2\,\mu$m of Seyfert 2s. To date, a detailed modelling of 
SEDs has been restricted  to individual galaxies 
(e.g., NGC~1068 Efstathiou, Hough, \& Young 1995; NGC~3281 
Simpson 1998; CenA Alexander et al. 1999; the 
Circinus galaxy Ruiz et al. 2000). 
An additional problem may exist when non-simultaneous observations
are used since some Seyfert 1.8 and 1.9 s are now found to 
show variability at  
near-infrared wavelengths 
over time scales of several months (Quillen et al. 2000a from 
{\it HST}/NICMOS observations at $1.6\,\mu$m).

Models of dusty tori with different geometries 
(e.g., Pier \& Krolik 1992; Granato \& Danese 1994;
Efstathiou \& Rowan-Robinson 1995;
Granato, Danese, \& Franceschini 1997) predict bright infrared 
emission, making the $L$-band (at $3.5\,\mu$m) and $M$-band (at
$4.8\,\mu$m) well suited to study the nature of the
non-stellar emission in Seyfert galaxies. Because of the strong
dependence of the infrared emission with the viewing angle and the
geometry of the obscuring material, infrared SEDs 
can be used to constrain the torus parameters in Seyfert galaxies. 
An additional advantage is 
that the stellar contribution is greatly reduced longwards of $2\,\mu$m,
even for Seyfert 2s (Alonso-Herrero et al. 1996). The 
use of observations at  
longer wavelengths $> 50\,\mu$m (e.g., IRAS and ISO measurements),
whilst useful in determining the bolometric emission, are problematic
when trying to separate the contributions from the 
central AGN and a possible circumnuclear starburst, because of the 
large apertures employed. 

In this paper we present near-infrared (from $1$ to $5\,\mu$m)
imaging observations of a small sample of Seyfert 1.9--2 galaxies which are 
complemented with {\it HST}/NICMOS and ISO imaging, covering a 
spectral range from 1 up to $15\,\mu$m. 
In a future paper (Quillen et al. 2001, in preparation) we will discuss the 
infrared properties and SEDs of the Seyfert 1.8--2s in the CfA sample. 
The paper is organized 
as follows, in Section~2 we describe the observations, in Section~3 
we determine the non-stellar emission, and in Section~4 we 
analyze the observed non-stellar energy distributions, non-stellar 
color-color diagrams, and compare them
with the outputs from torus models. In Section~5 we will show that the 
$5-10\,\mu$m fluxes are a good indicator for the AGN luminosity
for Compton-thin AGNs. Conclusions are presented in Section~6.

\begin{deluxetable}{lcccccc}
\footnotesize
\tablecaption{Log of the UKIRT and {\it HST}/NICMOS Observations.}
\tablehead{\colhead{Galaxy} & \colhead{UKIRT} & \colhead{Date} &
\colhead{NICMOS} & \colhead{Date}  & \colhead{var} & \colhead{timescale}\\
\colhead{(1)} & \colhead{(2)} & \colhead{(3)} &  \colhead{(4)} &
\colhead{(5)} & \colhead{(6)} & \colhead{(7)}}
\startdata
NGC~1052 & $JHKL^\prime M$ & Sep 1997 & NIC2 F160W & Sep 1998 & \nodata &
\nodata \\
NGC~1068 & $JHKL^\prime M$ & Sep 1997 & NIC2 F110W, F160W, F222M &
Feb 1998 & no & ???  \\
NGC~1097 & $JHKL^\prime M$ & Sep 1997 & \nodata & \nodata & 
\nodata & \nodata \\
NGC~2992 & $JHKL^\prime M$ & Apr 1998 & NIC2 F205W & Nov 1998 & \nodata &
\nodata \\
NGC~4968 & $JHKL^\prime M$ & Apr 1998 & NIC2 F160W & May 1998 & 
\nodata & \nodata \\
NGC~5252$^{\rm a}$ & $M$  &  Apr 1998 & NIC1 F110W, F160W & Mar 1998, 
Apr 1998 & no & 1 month\\ 
NGC~5506 & $JHKL^\prime M$ & Apr 1998 & NIC2 F160W, F205W & Apr 1998, 
Aug 1998 & \nodata & \nodata \\
NGC~7172 & $JHKL^\prime M$ & Sep 1997 & \nodata & \nodata & \nodata & 
\nodata\\
NGC~7217 & $JHKL^\prime M$ & Sep 1997 & NIC1 F110W, F160W & Aug 1997 &
\nodata & \nodata \\
Mkn~1 &    $JHKL^\prime$   & Sep 1997 & NIC1 F160W & Sep 1997 & 
\nodata & \nodata \\
Mkn~348$^{\rm a}$ & $M$    & Sep 1997 & \nodata& \nodata &
\nodata & \nodata \\
Mkn~533 &  $JHKL^\prime M$ & Sep 1997 & NIC1 F110W, F160W & Sep 1998, 
Sep 1997  & yes & 14 months \\ 
Mkn~573$^{\rm a}$ & $M$    & Sep 1997 & NIC1 F110W, F160W & Aug 1998 
& no & 12 months\\
MCG~$-$5-23-16 & $JHKL^\prime M$ & Apr 1998 & \nodata & \nodata &
\nodata & \nodata \\
\enddata
\tablecomments{$^{\rm a}$ $JHKL^\prime$ imaging can be found in 
Alonso-Herrero et al. (1998).}
\end{deluxetable}

\begin{deluxetable}{lccccc}
\tablewidth{17cm}
\tablecaption{UKIRT photometry through a 3\arcsec-diameter aperture
for the Seyfert 2 sample.}
\tablehead{\colhead{Galaxy} & \colhead{$J$} &
\colhead{$H$} & \colhead{$K$} & \colhead{$L^\prime$} & \colhead{$M$}}
\startdata
NGC~1052 & 11.40 & 10.66 & 10.39 & $9.77\pm0.02$ & $9.08\pm0.23$ \\
NGC~1068 & 10.42 & 9.08  &  7.54 & $4.53 \pm0.02$ & $3.24\pm0.02$\\
NGC~1097 & 11.95 & 10.99 & 10.75 & $10.19\pm0.17$ & $>10.50$ \\
NGC~2992 & 12.34 & 11.50 & 10.94 & $10.01\pm0.10$ & $9.15\pm0.30$\\  
NGC~4968 & 13.26 & 12.39 & 11.61 & $9.98\pm0.05$ & $8.65\pm 0.20$\\
NGC~5252$^{\rm a}$ & 13.27 & 12.56 & 12.03 & $10.58\pm0.04$ & $10.04\pm0.30$ \\
NGC~5506 & 12.09 & 10.46 &  9.14 & $7.11\pm0.03$ & $6.22\pm0.02$\\
NGC~7172 & 12.74 & 11.76 & 10.91 & $9.50\pm0.02$ & $8.56\pm0.05$\\
NGC~7217 & 11.99 & 11.27 & 10.96 & $10.88\pm 0.15$ & $> 10.60$ \\
Mkn~1    & 14.12 & 13.33 & 12.89 & $11.26\pm0.10$ & \nodata\\
Mkn~348$^{\rm a}$ & 13.52 & 12.72 & 12.17 & $10.50\pm0.04$ & $9.06\pm 0.10$ \\
Mkn~533 &  13.38 & 12.28 & 11.21 & $9.14\pm 0.02$ & $7.95\pm 0.08$ \\
Mkn~573$^{\rm a}$ & 12.87 & 12.27 & 11.78 & $10.14\pm0.04$ & $8.98\pm 0.30$ \\
MCG~$-$5-23-16 & 11.94 & 11.01 & 10.22 & $8.51\pm0.02$ & $7.67\pm 0.06$\\ 
\enddata
\tablecomments{$^{\rm a}$$JHKL^\prime$-band photometry from 
Alonso-Herrero et al. (1998).\\
The quoted errors for the $L^\prime$ and $M$-band 
photometry account only for the sky subtraction uncertainties, except for 
the Alonso-Herrero et al. (1998) $L^\prime$-band 
photometry which are the photometric errors.\\
The $M$-band limits for NGC~1097 and NGC~7217 are 3$\sigma$.}
\end{deluxetable}

\begin{deluxetable}{lccc}
\tablewidth{12cm}
\tablecaption{Mid-infrared ISO fluxes.}
\tablehead{\colhead{Galaxy} & \colhead{$\lambda$} &
\colhead{flux} & \colhead{reference}\\
     &    & [mJy] & \\
(1) & (2) & (3) & (4) }
\startdata

NGC~1068 & $5.9\,\mu$m & 8660 & Rigopoulou et al. (1999)\\ 
         & $7.7\,\mu$m & 14500 & Rigopoulou et al. (1999)\\ 
NGC~3227 & $6.75\,\mu$m & 294, 249 & Clavel et al. (2000)\\
         & $9.63\,\mu$m & 382, 372 & Clavel et al. (2000)\\
         & $16\,\mu$m & 1220 & P\'erez Garc\'{\i}a \& Rodr\'{\i}guez
Espinosa (2000)\\
NGC~3281 & $11.4\,\mu$m & 530      & this work \\
NGC~4151 & $5.9\,\mu$m  & 882 & Rigopoulou et al. (1999)\\
         & $7.7\,\mu$m  & 1090 & Rigopoulou et al. (1999)\\
         & $16\,\mu$m & 4120 & P\'erez Garc\'{\i}a \& Rodr\'{\i}guez
Espinosa (2000)\\
NGC~5252 & $6.75\,\mu$m & 14.8 & this work\\   
         & $9.63\,\mu$m & 27.1 & this work \\   
         & $11.4\,\mu$m & 34.6 & this work\\   
         & $15\,\mu$m   & 44.8 & this work\\   
NGC~5506 & $5.9\,\mu$m  & 646  & Rigopoulou et al. (1999)\\ 
         & $7.7\,\mu$m  & 713  & Rigopoulou et al. (1999)\\ 
         & $11.4\,\mu$m  & 890  & this work\\
NGC~5548 & $6.75\,\mu$m & 170 & Clavel et al. (2000)\\
         & $9.63\,\mu$m & 270 & Clavel et al. (2000)\\
         & $16\,\mu$m & 440 & P\'erez Garc\'{\i}a \& Rodr\'{\i}guez
Espinosa (2000)\\
NGC~7172 & $6.75\,\mu$m & 160 & this work\\
         & $15\,\mu$m   & 240 & this work\\
NGC~7469 & $5.9\,\mu$m  & 270 & Rigopoulou et al. (1999)\\
         & $6.75\,\mu$m & 433 & this work\\
         & $7.7\,\mu$m  & 539 & Rigopoulou et al. (1999)\\
         & $15\,\mu$m   & 1290 & this work\\ 
Mkn~1    & $5.9\,\mu$m  & 20.9 & Rigopoulou et al. (1999)\\
         & $7.7\,\mu$m  & 88.3 & Rigopoulou et al. (1999)\\
Mkn~348  & $6.75\,\mu$m  & 65   & this work \\
         & $15\,\mu$m    & 270  & this work \\
Mkn~533  & $6.75\,\mu$m & 259, 213 & Clavel et al. (2000)\\   
         & $9.63\,\mu$m & 345 &  Clavel et al. (2000)\\ 
IC~4329A & $6.75\,\mu$m & 591, 477  & Clavel et al. (2000)\\ 
         & $9.63\,\mu$m & 890, 739  & Clavel et al. (2000)\\ 
\enddata
\end{deluxetable}

\section{OBSERVATIONS}
\subsection{UKIRT Observations}
We obtained $J$ ($\lambda_{\rm c} = 1.25\,\mu$m), $H$ 
($\lambda_{\rm c} = 1.65\,\mu$m), $K$
($\lambda_{\rm c} = 2.20\,\mu$m), $L^\prime$
($\lambda_{\rm c} = 3.80\,\mu$m) and $M$ 
($\lambda_{\rm c} = 4.80\,\mu$m) imaging 
of a sample of fourteen Seyfert 1.9 and Seyfert 2 
galaxies (see Table~1) with the infrared camera 
IRCAM3 on the 3.9-m United Kingdom 
Infrared Telescope (UKIRT) during two observing runs in  
September 1997 and April 1998.  The size 
of the $J$, $H$ and $K$ images is $256 \times 256\,$pixels with pixel
size of 0\farcs143 pixel$^{-1}$, 
whereas the size of the $L^{\prime}M$  images is $64 \times 64\,$pixels,
with pixel size 0\farcs286 pixel$^{-1}$.

Standard reduction procedures were applied. Conditions were photometric 
during both runs,  
so standard star observations from the Elias et al. (1982) list 
(for $JHKL^\prime$) and from the UKIRT list of photometric standards 
(for $M$) were used to perform the photometric calibration of the 
images. The major source of uncertainty in near-infrared observations 
comes from the background subtraction, especially at the longer 
wavelengths ($L^\prime$ and $M$). 
Typical errors from the photometric calibration are 0.09, 0.04, 0.08, 
0.07  and 0.10\,mag in $J$, $H$, $K$, $L^{\prime}$ and $M$, 
respectively. The full width half maximum (FWHM) seeing  
was measured from the standard stars 
$0.6-0.7$\arcsec \ at $K$ band for both runs. 

Standard synthetic aperture photometry was performed on all the 
images using a 3\arcsec-diameter 
circular aperture (Table~2). The errors associated with the sky subtraction
at $L$ and $M$ are given together with the aperture photometry in 
the last two columns of Table~2.

\subsection{{\it HST}/NICMOS Observations}
We searched the {\it HST} archive for broad-band
NICMOS observations of our sample of 
galaxies. In Table~1 columns~(4) and (5) we list the camera and filter, and   
the observation date respectively. In columns~(6) and (7) we indicate if
variability has been found, and the time scale from Quillen et al. 
(2000a). 
The plate scales for cameras NIC1 and NIC2 are
0.045"\,pixel$^{-1}$ and 0.076"\,pixel$^{-1}$ respectively.

The images were reduced with routines from  the package
NicRed (McLeod 1997).  The main steps in the data reduction involve 
subtraction of the first readout, dark current subtraction on a 
readout-by-readout basis, correction for linearity and cosmic
ray rejection (using fullfit), and flatfielding. 
Darks with sample sequences and exposure times
corresponding to those of our observations were obtained from other 
programs close in time to ours. Usually between 10 and 20 darks were 
averaged together (after the subtraction of the first readout) for a 
given sample sequence. Flatfield images were constructed from  on-orbit data. 

The photometric calibration of the NICMOS images was performed 
using the conversion factors based on measurements of the standard star P330-E
(Marcia Rieke 1999, private communication). Comparisons of the ground-based 
$H$-band aperture photometry with the NICMOS F160W filter show that 
differences are always less than $15\%$, whereas the comparison
of ground-based $K$-band photometry with that of the NICMOS F222M and F205W
filters agrees within $25\%$.

\subsection{Comparison sample of Seyfert 1s}
As a comparison sample, we also searched the {\it HST} archive for 
Seyfert 1--1.5 galaxies with $1-2.2\,\mu$m NICMOS observations as well as with 
ground-based $L$ (or $L^\prime$) and $M$ small aperture photometry 
and ISO observations. 

\subsection{ISO Observations}
We searched the literature  
for available mid-infrared ISO (from $5\,\mu$m up to $15\,\mu$m) 
continuum fluxes. In addition, for those galaxies with unpublished
data, we obtained images from the ISO archive. The photometric 
calibration was that of the pipeline
for which the typical errors are $\pm 15\%$. 
The ISO fluxes were measured using 12\arcsec-diameter apertures
and aperture corrections measured from model PSFs.
This procedure is very similar to that employed by Clavel et
al (2000). NGC~5506 and NGC~7469
looked like point sources so contamination
by nearby star formation was not likely to be a large problem.
For the brightest source (NGC~5506) some additional problems
may be caused by saturating the detector. The wavelengths, fluxes and 
references for the ISO data for the 
galaxies in our sample of Seyfert 2s and 
Seyfert 1s  are given in Table~3.

\section{DETERMINATION OF THE NON-STELLAR FLUXES}

\subsection{The $1-2.2\,\mu$m spectral range}

As mentioned in the introduction it is essential to obtain 
very accurate estimates of the non-stellar emission to 
understand and model the SEDs of Seyfert 2s. The stellar 
component may dominate most of the nuclear flux for wavelengths 
up to $2\,\mu$m in Seyfert 2s (Alonso-Herrero et al. 1996).  
The relatively small field of view of the ground-based 
$JHK$ ($36\arcsec \times 36\arcsec$) and NICMOS images
($19\arcsec \times 19\arcsec$ and $11\arcsec \times 11\arcsec$
for NIC2 and NIC1 observations respectively) prompted us 
to try a nucleus + bulge decomposition of the observed radial surface 
brightness profiles. We used 
a similar method to that employed in Alonso-Herrero et al. (1996). 
For the ground-based UKIRT observations the nuclear component was 
assumed to be a delta function convolved a Gaussian profile to account for 
the seeing effects (see Alonso-Herrero et al. 1996 for more details). For the
NICMOS observations, the nuclear component was represented with 
a surface brightness profile of a point spread function (PSF) 
generated with the TinyTim software 
(Krist et al. 1998). The bulge component was represented
with a de Vaucouleurs $r^{\frac{1}{4}}$ profile. The surface brightness
profiles were extracted out to radial distances of 5.7\arcsec, 
4.6\arcsec, and 2.6\arcsec
\ for the ground-based data, NIC2 and NIC1 data respectively.

\begin{deluxetable}{lccccccccc}
\tablewidth{17cm}
\tablecaption{Non-stellar fluxes and non stellar contribution 
within a $3\arcsec$-diameter aperture for the Seyfert 2 sample.}
\tablehead{\colhead{Galaxy} & \colhead{$f(J)$} &
\colhead{$f(H)$} & \colhead{$f(K)$} & \colhead{$f(L^\prime$)} & 
\colhead{$f(M)$} & \colhead{$J$} & \colhead{$H$} &
\colhead{$K$} & \colhead{$L^\prime$}\\
 & [mJy]  &[mJy]  &[mJy]  &[mJy]  &[mJy] &
 [\%] & [\%] & [\%] & [\%] \\
(1) & (2) & (3) & (4) & (5) & (6) & (7) & (8) & (9) & (10)}

\startdata
NGC~1052 & no  & no    & $<0.9$   & $<20.2$   & 38.0    & no & no  (0) &
$<2$ & $<65$\\ 
NGC~1068 & 9.8 & 97.6 & 449.6  & 3690.8 & 8245.0  &  17 (9) & 47 (41) & 
69 (71) & 95 \\
NGC~2992 & no   & $<1$  & 2.8   & 22.7   & 35.7    & no & no & 10 (23) & 91 \\ 
NGC~4968 & no   & 0.6   & 3.7   & 23.2   & 56.5    & no &  5 (7) & 25 & 
90 \\
NGC~5252 & no   & 0.7   & 1.0   & 11.1   & 15.7    & no & 4  (7) & 10  & 75 \\ 
NGC~5506 & 13.8 & 59.0 & 120.4  & 340.1  & 530.0   & 59 & 79 (89) & 64 (83) 
& 95 \\ 
NGC~7172 & no   & $<0.4$ & 3.4  & 30.0   & 61.4    & no & $<2$ & 12 & 75 \\
Mkn~1    & $<0.07$ & 0.15 & 0.8 & 6.4    & \nodata & $<2$  & 3 (7) & 18 & 
80 \\ 
Mkn~348  &  $<1$  & 0.6  & 2.7   & 14.5   & 38.7    & $<2$ & 7 & 30 & 91 \\
Mkn~533  & 1.0  & 5.0   & 12.3  & 53.0   & 108.6   & 23 (14) & 31 (40) & 57 & 
95 \\ 
Mkn~573  & 0.2  & 0.6   & 3.2   & 18.8   & 41.3    &  (2) & (6) & 25  &85\\
MCG~$-$5-23-16 & 1.1 & 3.7 & 10.7  & 79.5   & 139.4   &   4  &  9  & 20 &  80 \\ 
\enddata
\tablecomments{The zero points used for $JHKL^\prime M$ are:
$1600\,$Jy, $1020\,$Jy, $657\,$Jy, $252\,$Jy and $163\,$Jy respectively. The 
3\arcsec-diameter $M$-band fluxes are assumed to be all non-stellar.
``No'' means that no unresolved component was detected.}
\end{deluxetable}

\begin{deluxetable}{lcccccccc}
\tablewidth{17cm}
\tablecaption{Non-stellar fluxes and non stellar contribution 
within a $3\arcsec$-diameter aperture for the Seyfert 1 sample.}
\tablehead{\colhead{Galaxy} & \colhead{$f({\rm F110W})$} &
\colhead{$f({\rm F160W})$} & \colhead{$f({\rm F222M})$} & \colhead{$f(L$)} & 
\colhead{$f(M)$} & \colhead{$J$} & \colhead{$H$} &
\colhead{$K$}\\
 & [mJy]  &[mJy]  &[mJy]  &[mJy]  &[mJy] &
 [\%] & [\%] & [\%]  \\
(1) & (2) & (3) & (4) & (5) & (6) & (7) & (8) & (9)}
\startdata
NGC~3227 & \nodata & 10.6 & 22.6 & $78.3\pm4.4$ & $72\pm27$ & \nodata &
21 & 32 \nl 
NGC~4151 & 69.0 & 103.6 & 177.5   & $325\pm10$   & $449\pm34$ & 82 &
81 & 77\\
NGC~5548 & \nodata & 15.0 & 31.6 & $98.6\pm3.7$ & $100\pm21$ & \nodata &
57 & 59 \\
NGC~7469 & 16.2    & 39.0 & 67.8 & $159\pm5$ & $259\pm33$ & 45 &
57 & 63\\
IC~4329A & \nodata & 49.9 & 101.8 & $210\pm5$ & $170\pm20$ & \nodata &
71 & 75 \\
\enddata
\tablecomments{The $L$ and $M$ fluxes are from Ward et al. 
(1987), and are assumed to be all non-stellar in origin.}
\end{deluxetable}

In general, we performed a nucleus + bulge deconvolution to the radial
surface brightness profiles, as advocated by Simpson (1994). For
Mkn~348, however, only lower angular-resolution data were available
(from Alonso-Herrero et al.\ 1998), and we opted for the method of
Simpson (1998), which involves scaling and subtracting the $J$-band
image to the $H$ and $K$ images, according to the counts in a
circumnuclear annulus. This method assumes that the non-stellar
contribution at $J$ is negligible and, like Simpson (1998), we
confirmed this by extrapolating the stellar $HKL'M$ fluxes to the
$J$-band.

After performing the nucleus + bulge
deconvolution we measured the unresolved flux, which will be assumed to 
be non-stellar in origin, and its contribution within a 
3-arcsec diameter aperture. The non-stellar fractions within
this aperture are given in columns (7), (8) and (9) for $JHK$ 
respectively for the ground-based data, and in parenthesis for the
NICMOS data when available.  The non-stellar fraction 
within a 3-arcsec aperture (which corresponds to projected
angular sizes of between 200\,pc and 1.5\,kpc for our 
sample of galaxies) gives an estimate
of the dominance or otherwise of the non-stellar component at 
a given wavelength. In Table~4, columns (2), (3) and  
(4) we give the  $JHK$ non-stellar fluxes using the NICMOS non-stellar 
fractions (when available) and the ground-based photometry. Note that 
we did not detect an unresolved component at $1-2.2\,\mu$m
in NGC~1097 and NGC~7217, and therefore these two galaxies 
are excluded from the following analysis.
We performed a similar analysis on the NICMOS images as with the 
Seyfert 2s to derive the non-stellar fluxes from $1$ to $2.2\,\mu$m
(see Table~5).

As discussed in Alonso-Herrero 
et al. (1996) the main source of uncertainty in the 
determination of the unresolved fluxes from 
the ground-based data is the fitted value to the 
seeing. Because of the relatively small field of view of the 
images, no stars are present in the images. However, the fitted values 
of the seeing from the profile deconvolution 
are very similar to the values of the FWHM measured from the standard
stars observed close in time. On the other hand, the advantage of
using  NICMOS images is that the NIC1 and NIC2 PSF is 
very stable, is very well sampled and can be 
well modeled with the TinyTim software. Therefore the 
estimated non-stellar fluxes from NICMOS profiles are less affected 
by uncertainties. An upper limit to the errors
associated with the non-stellar flux determination can be obtained by
comparing the non-stellar fractions derived from the ground-based and 
the NICMOS images. The differences are never greater than 45\%, and tend 
to increase for shorter wavelengths, where the  non-stellar contributions
are smaller. 

Our estimates are in relatively good agreement (to within the uncertainties
given above) for those galaxies in common 
with Zitelli et al. (1993), 
Kotilainen et al. (1992), Alonso-Herrero et al. (1996) and 
Quillen et al. (2000b). 
We confirm the findings of  Alonso-Herrero et al. (1996), 
that the 1 to $2.2\,\mu$m continuum of some 
Seyfert 2 galaxies is dominated by stellar emission. However, the stellar 
contribution decreases significantly at longer wavelengths. 

\subsection{The $3-15\,\mu$m spectral range}

At the $L^{\prime}$-band and especially at  
the $M$-band the host galaxy does 
not make a strong contribution. We used the $L^\prime$ non-stellar fractions
derived by Alonso-Herrero et al. (1998) for the galaxies in common with 
our study. For the rest of the sample we estimated 
the non-stellar fractions at $L^\prime$ scaling the peak of the PSF surface 
brightness profile to that of the galaxy, and assumed that the 
PSF integrated flux represents the unresolved
component (see Alonso-Herrero et al. 1998 for
more details). An alternative way of estimating
the stellar contribution in the $L^\prime$ can 
be done by using the typical colors $K-L^\prime$ of 
a normal galaxy (Willner et al. 1984, and assuming 
a color transformation $L-L^\prime = 0.2$, Ward et 
al. 1982) and the stellar fractions at the $K$-band. We find
that the stellar contribution within a 3\arcsec-diameter aperture 
at the $L^\prime$ are always less than $25\%$ (except  
NGC~1052 for which we find a $65\%$ of the total 
3\arcsec emission is stellar emission) and 
are in good agreement with the values estimated from the imaging
data. The same procedure was applied to the $M$-band fluxes using
the $K-M$ color of a normal galaxy (Rieke \& Lebofsky 
1978), and estimated that the 
stellar contribution at $4.8\,\mu$m within a 3\arcsec-diameter
aperture is always less than $10\%$. 
This is supported by the compact appearance of all the observed 
sources at this wavelength.
In Table~4, columns (5) and (6) we list the non-stellar fluxes 
at $L^\prime$ and $M$. 
 
The small aperture $L$ and $M$ fluxes for the 
Seyfert 1 sample (compiled by Ward et al. 
1987) are assumed to be all non-stellar in origin. The
non-stellar fluxes along with the non-stellar fractions within a 
3-arcsec diameter aperture are given in Table~5. 
Note that the non-stellar $1.2-2.2\,\mu$m 
fluxes are not simultaneous with the $3.5-4.8\,\mu$m observations, 
which may add some uncertainty to the observed SEDs.

\begin{deluxetable}{lcc}
\tablewidth{8cm}
\tablecaption{Viewing angles, UV optical depths, and 
equivalent visual extinctions for the models.}
\tablehead{\colhead{$\theta_{\rm v}$}  & \colhead{$\tau_{\rm UV}$} &
\colhead{$A_V$}\\
\colhead{(\arcdeg)} & \colhead{(mag)} & \colhead{(mag)}}
\startdata
90   & 1200 & 221 \\
84   & 1197 & 221 \\
79   & 1018 & 188 \\
73   & 914  & 169 \\
64   & 812  & 150 \\
50   & 700  & 129 \\
44   & 658  & 121 \\
40   & 620  & 114 \\
36   & 558  & 103  \\
30   & 263  & 49   \\
30   & 50 (24)   & 9 (4) \\
25   & 2.5 (0)  & 0.5 (0) \\
20   & 2.5 (0)  & 0.5 (0) \\
10   & 2.5 (0)  & 0.5 (0) \\
0    & 2.5 (0)  & 0.5 (0) \\
\enddata
\tablecomments{In columns~(2) and (3) in brackets
are the values for the $T=1500\,$K model.}
\end{deluxetable}

\section{THE NON-STELLAR INFRARED EMISSION OF SEYFERT GALAXIES} 
Because the non-stellar contribution to the 
near-infrared continuum emission from Seyfert 1s usually dominates 
over the stellar emission, unlike the situation for Seyfert 2s 
(see e.g., Edelson \& Malkan  1986; Kotilainen et al. 1992 and Alonso-Herrero 
et al. 1996), the SEDs of the non-stellar component is 
better defined in the Seyfert 1s. 
It is agreed that the infrared (up to $100\,\mu$m) SEDs of 
Seyfert 1s  are well represented with a power law ($S_\nu
\propto \nu^{-\alpha}$) with a canonical spectral 
index $\alpha \simeq 1$ (e.g., Edelson \& Malkan 1986; 
Ward et al. 1987; Edelson, Malkan, \& Rieke 1987; 
Granato \& Danese 1994; Clavel et al. 2000). Similar spectral indices have
been derived from the infrared SEDs of quasars (Edelson 1986; Neugebauer 
et al. 1987; Polletta et al. 2000; Haas et al. 2000). 
Fadda et al. (1998) have compiled infrared data from the literature
for a sample of Seyfert 1s and Seyfert 2s, and found that the 
non-stellar SEDs of Seyfert 2s are always steeper than those of 
Seyfert 1s, in line with the finding that the
$1-2.2\,\mu$m non-stellar fluxes of Seyfert 2s   are 
significantly smaller than those of Seyfert 1s. However, 
the work of  Fadda et al. (1998) and that of 
Murayama, Mouri, \& Taniguchi (2000) suffer from a number of 
limitations when attempting to place constrains on the torus geometry based on 
the non-stellar infrared SEDs or colors. These works make use 
of non-simultaneous observations, observations from different sources, 
uncertain non-stellar fluxes,  and relatively large aperture $N$- 
and $Q$-band or IRAS fluxes which may be contaminated by other 
contributions. Moreover, in order to understand 
the nature of the non-stellar SEDs of Seyfert 2 galaxies and in 
turn to constrain the torus geometry and derive the infrared 
extinctions to the central regions and their relation 
with the hard X-ray absorption columns, one cannot average the non-stellar 
SEDs of Seyfert 2s (or Seyfert 1s). 

\begin{figure*}
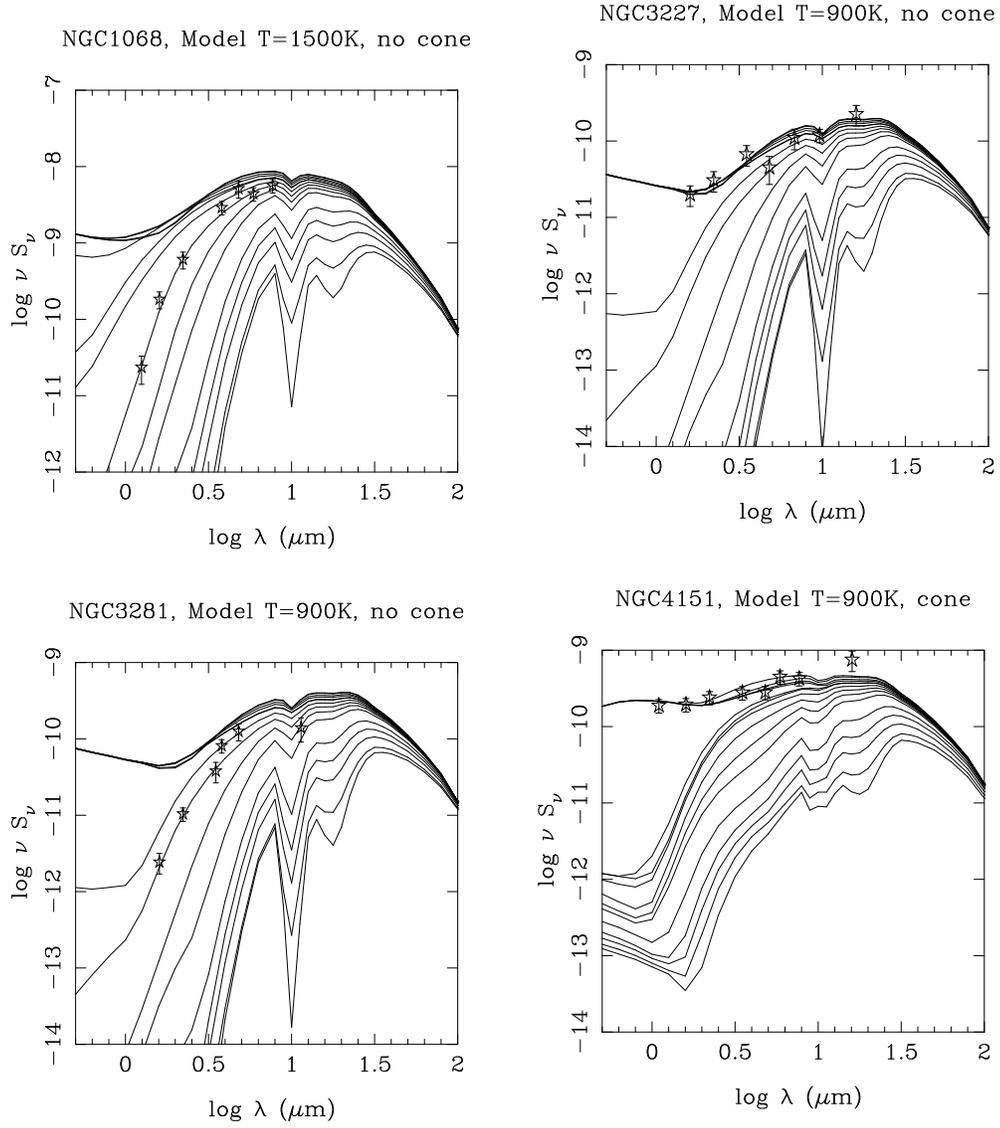

\figurenum{1}
\plotfiddle{alonso.fig1a.ps}{425pt}{270}{50}{50}{-200}{525}
\plotfiddle{alonso.fig1b.ps}{425pt}{270}{50}{50}{0}{970}
\plotfiddle{alonso.fig1c.ps}{425pt}{270}{50}{50}{-200}{1180}
\plotfiddle{alonso.fig1d.ps}{425pt}{270}{50}{50}{0}{1620}
\vspace{-45cm}
\caption{Observed SEDs and best-fitting models for 
the sample of Seyfert galaxies. The viewing angles
($\theta_{\rm v}$) for the model outputs increase 
from top to bottom, and are given in Table~6. For NGC~7469 
we show the two best fitting models (cone and coneless).}
\end{figure*}

\begin{figure*}
\figurenum{1}
\plotfiddle{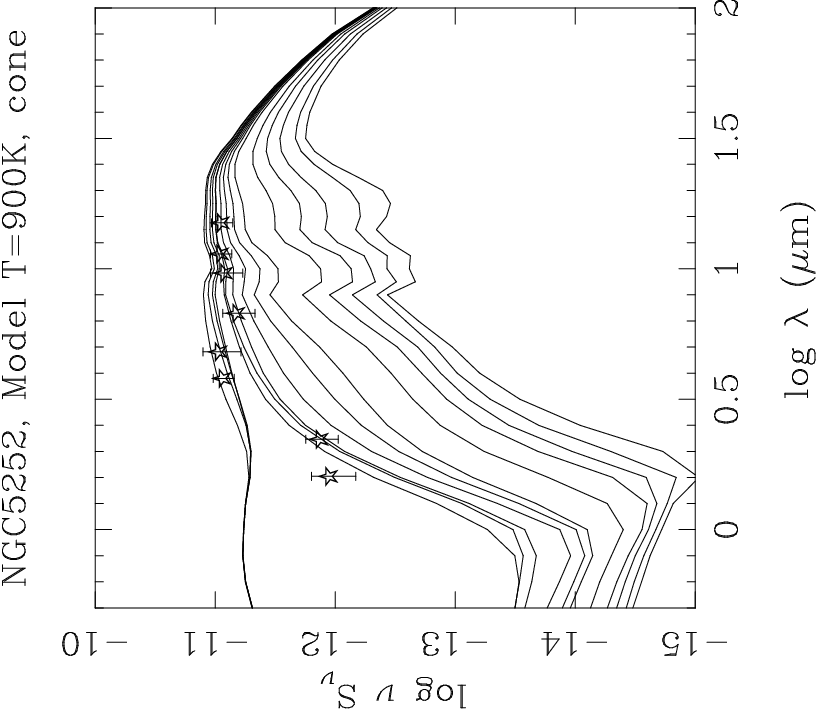}{425pt}{270}{50}{50}{-200}{525}
\plotfiddle{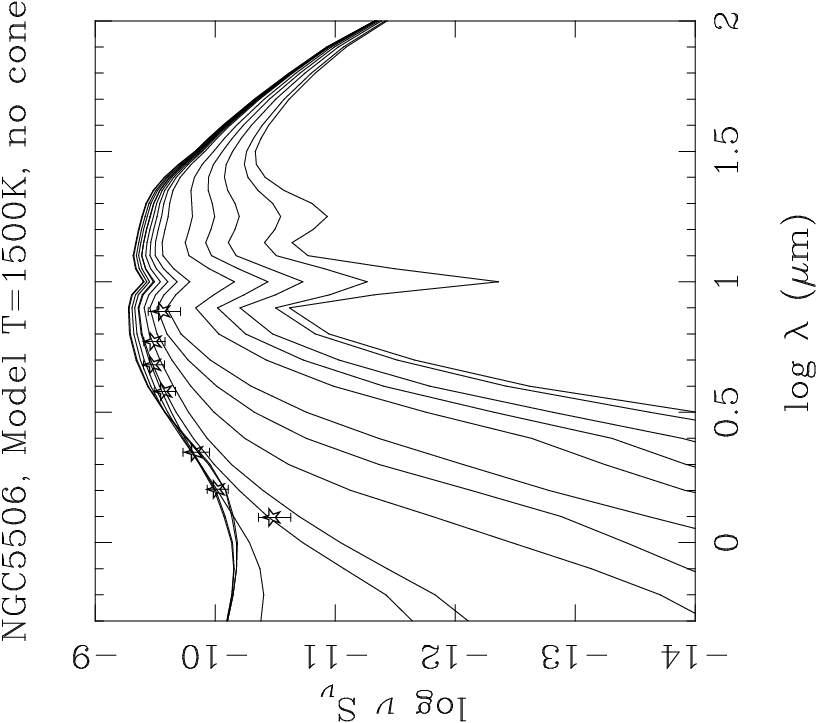}{425pt}{270}{50}{50}{0}{960}
\plotfiddle{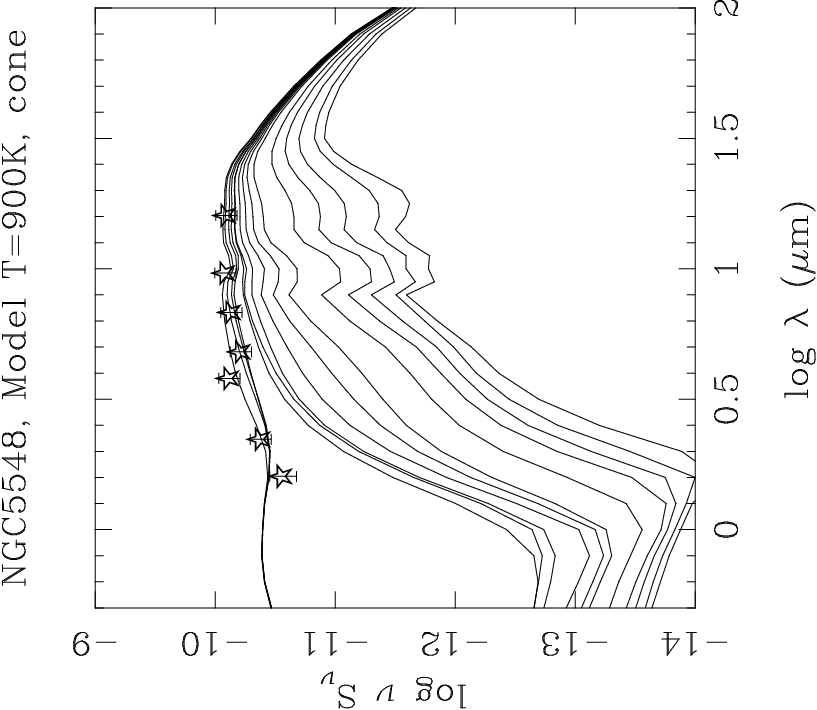}{425pt}{270}{50}{50}{-200}{1180}
\plotfiddle{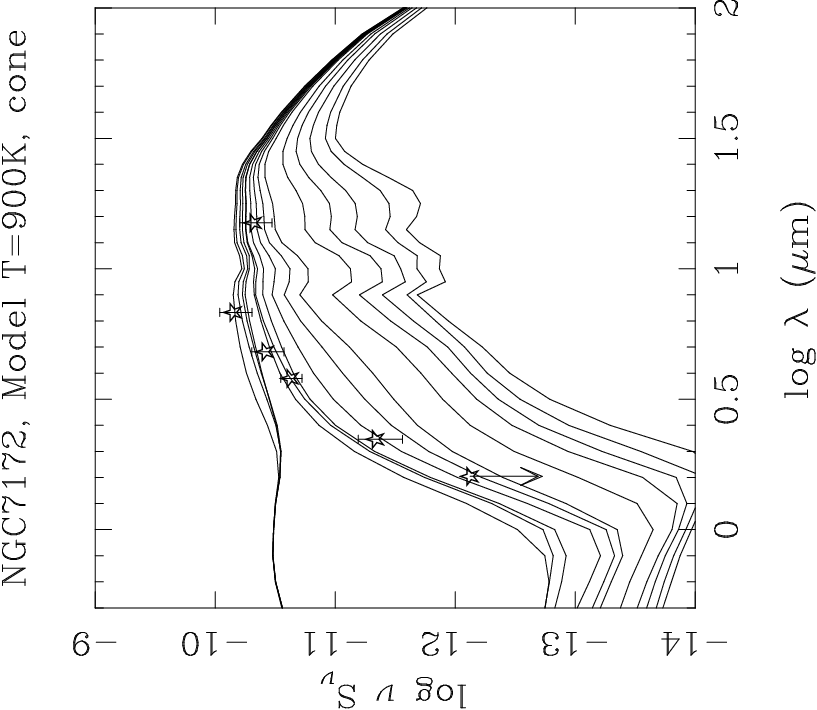}{425pt}{270}{50}{50}{0}{1620}
\plotfiddle{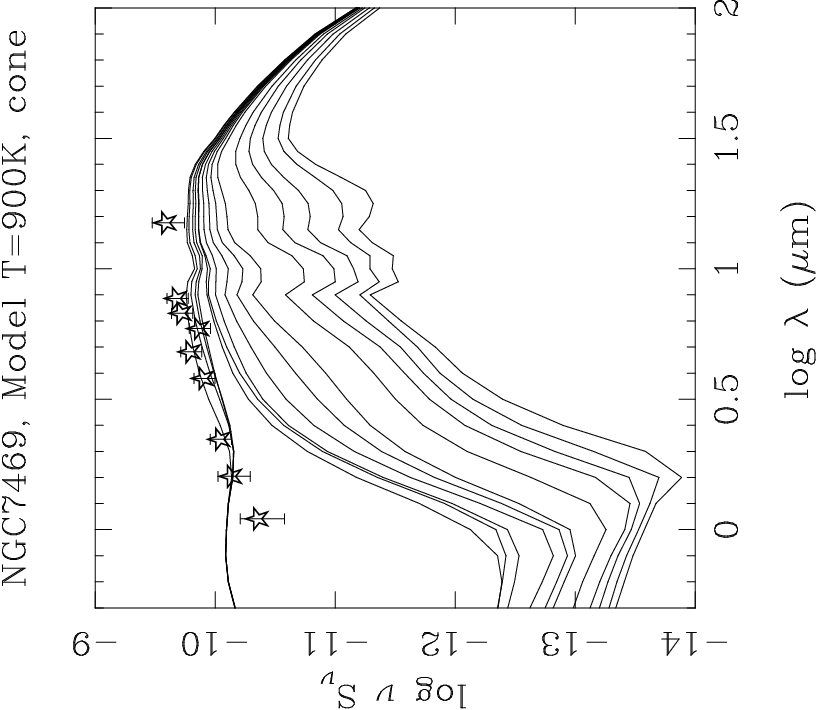}{425pt}{270}{50}{50}{-200}{1820}
\plotfiddle{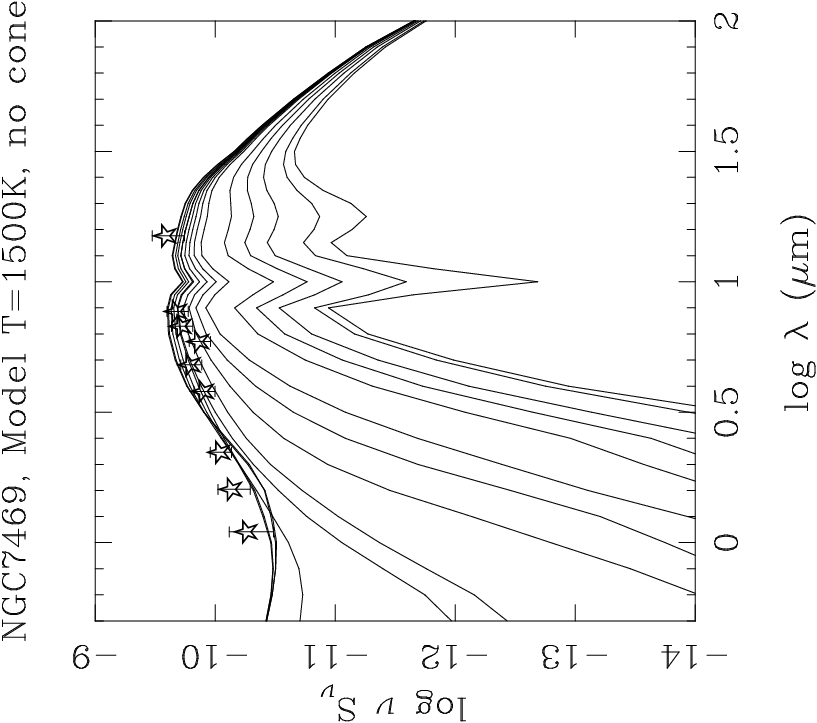}{425pt}{270}{50}{50}{0}{2250}
\vspace{-68cm}
\caption{Continued.}
\end{figure*}

\begin{figure*}
\figurenum{1}
\plotfiddle{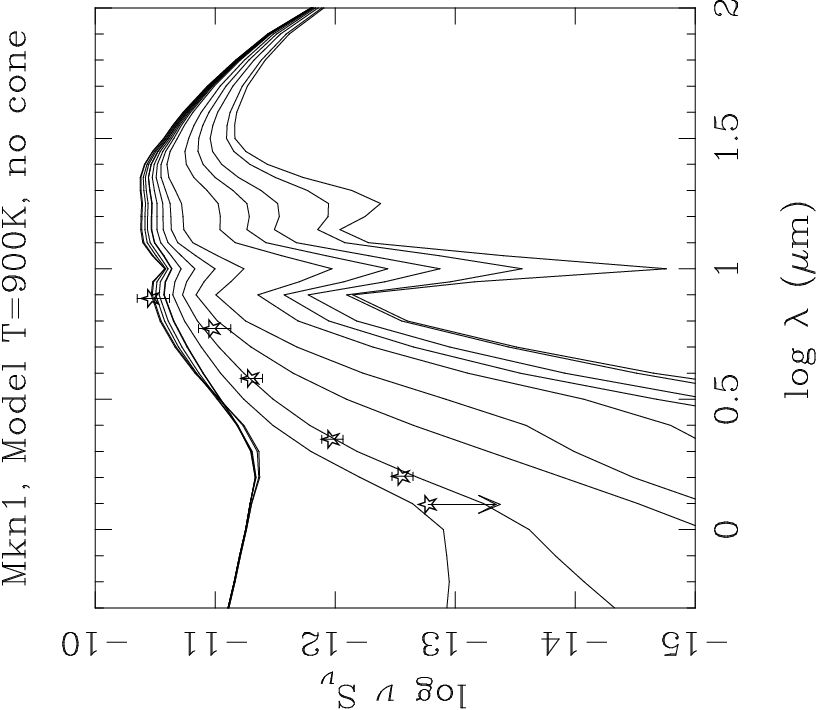}{425pt}{270}{50}{50}{-200}{525}
\plotfiddle{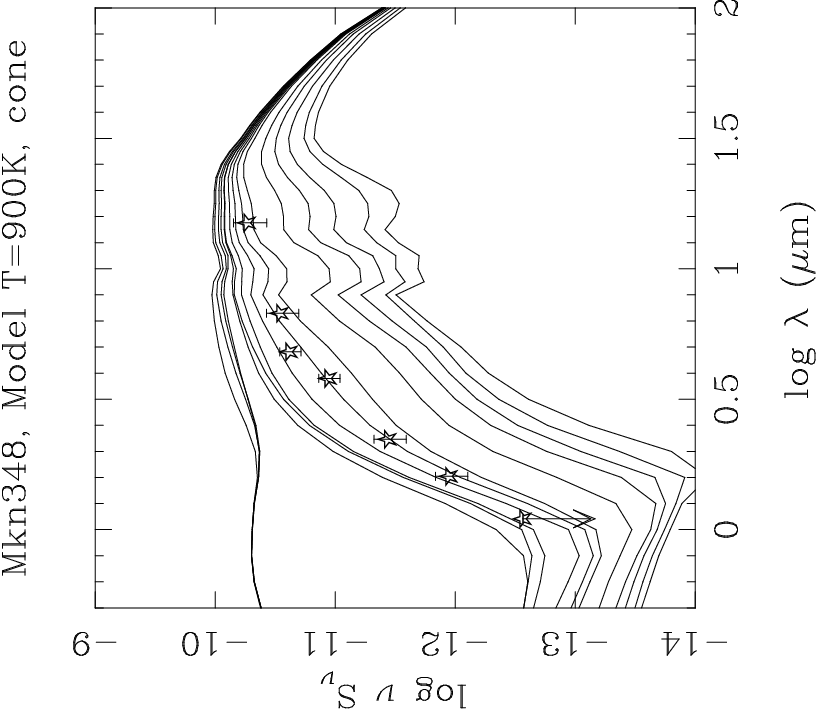}{425pt}{270}{50}{50}{0}{960}
\plotfiddle{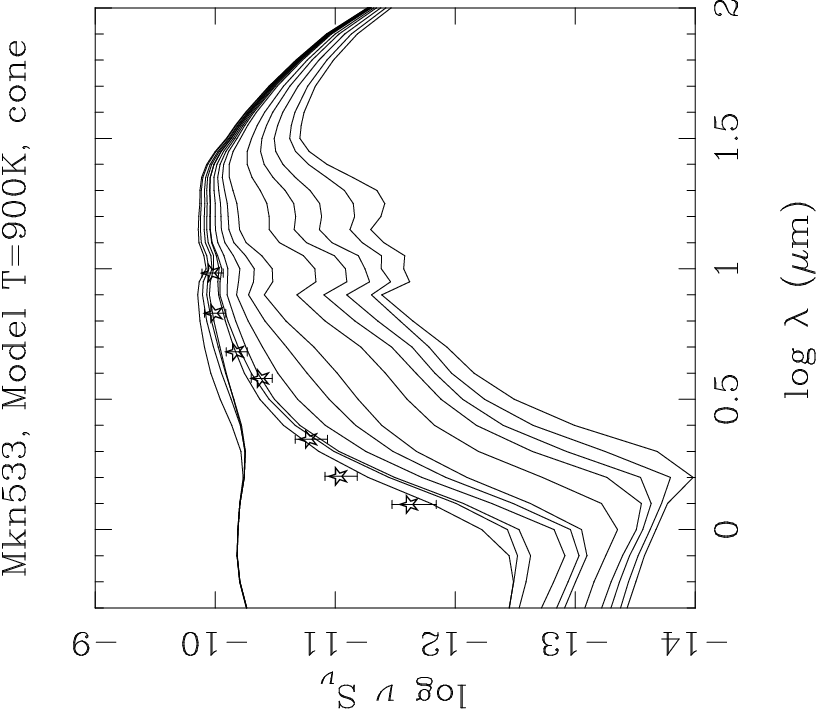}{425pt}{270}{50}{50}{-200}{1180}
\plotfiddle{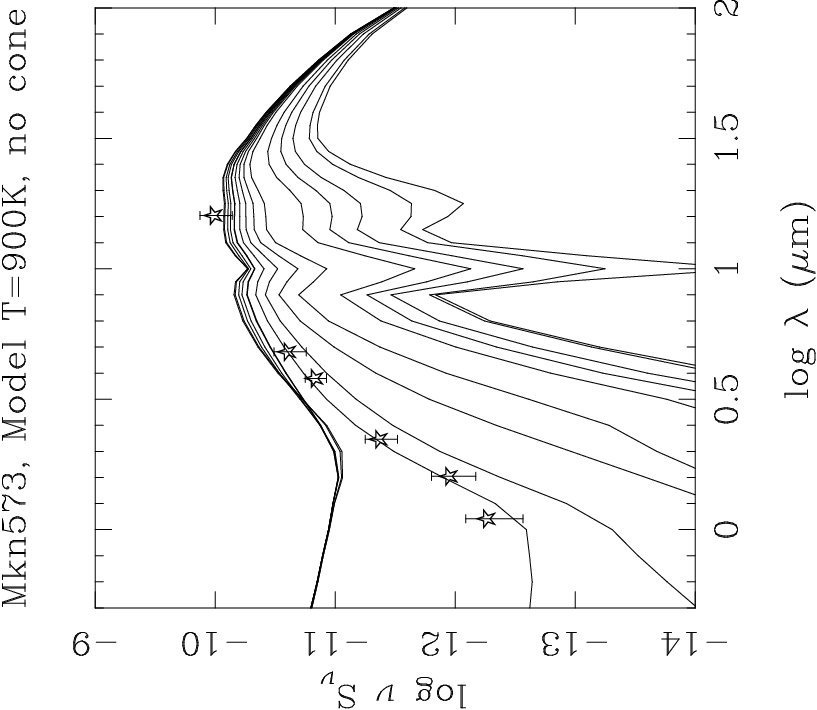}{425pt}{270}{50}{50}{0}{1620}
\plotfiddle{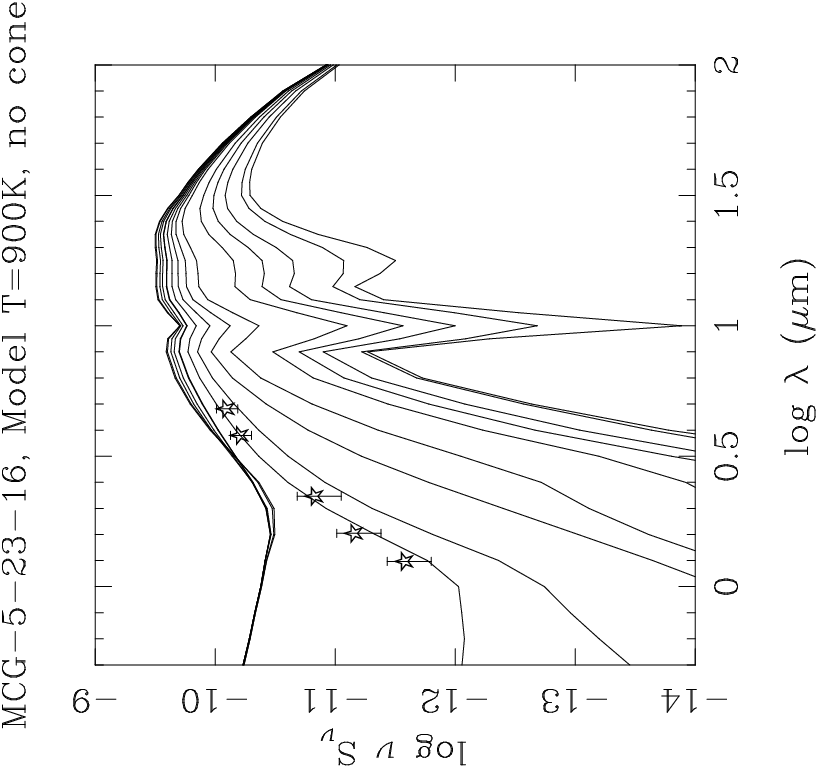}{425pt}{270}{50}{50}{-200}{1820}
\plotfiddle{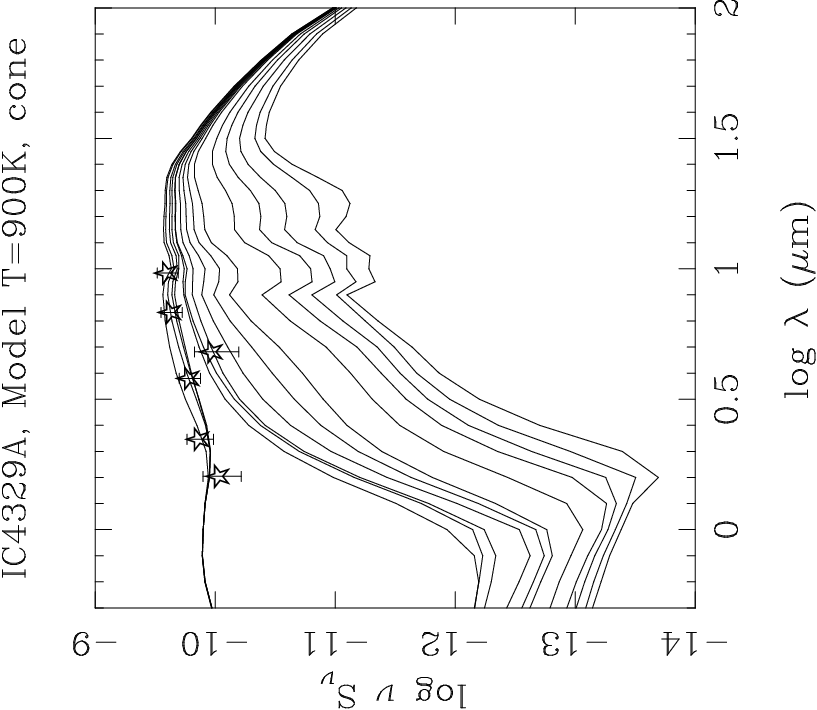}{425pt}{270}{50}{50}{0}{2250}
\vspace{-68cm}
\caption{Continued.}
\end{figure*}

An additional problem often present in past statistical studies of 
SEDs is 
the grouping of galaxies into the Seyfert 1 and Seyfert 2 categories. A 
galaxy may be classified as a Seyfert 2 when it is observed in
the optical (that is, no broad lines), but it may still show 
broad lines in the infrared, where the extinction is greatly 
reduced (see for instance, 
Ruiz, Rieke, \& Schmidt 1994; Goodrich, Veilleux, \& Hill 1994;
Veilleux, Goodrich, \& Hill 1997). 
If using the latter observations, one would classify it as a Seyfert 1.8 or 
Seyfert 1.9. A better approach to the understanding of the infrared SED of
Seyfert galaxies is to study each galaxy separately and determine whether 
a single torus geometry is able to fit most of the galaxies. 
Although in our study we do not claim a complete or unbiased
sample of Seyfert 1s and 2s, because we have estimated accurate
non-stellar SEDs, we can compare these with current torus
models and see whether there is a general agreement across 
the range of properties present in this sample. 
We will make use of Efstathiou \& 
Rowan-Robinson (1995) models, and discuss other models available 
in the literature.  In addition to the galaxies presented in this 
work, we will fit the non-stellar infrared SED of NGC~3281 
(non-stellar fluxes from Simpson 1998).

\begin{deluxetable}{lcccc}
\tablewidth{14cm}
\tablecaption{Infrared non-stellar SED modelling.}
\tablehead{\colhead{Galaxy} & \colhead{n} & \colhead{$\chi^2/\nu$} &
 \colhead{$\theta_{\rm v}$}  & \colhead{model}}
\startdata
NGC~1068 & 7 & $1.0-1.6$ & 40\arcdeg & Cone, $T=900\,$K and 
No cone, $T=1500\,$K\\
NGC~3227 & 7 & 1.0 & 30\arcdeg & No cone, $T=900\,$K\\
NGC~3281 & 6 & $1.8-2.2$ & 40\arcdeg & Cone and No cone, $T=900\,$K \\
NGC~4151 & 8 & 0.8 & 10\arcdeg & Cone, $T=900\,$K\\
NGC~5252 & 8 & 1.6 & 30\arcdeg & Cone, $T=900\,$K \\
NGC~5506 & 7 & 1.3   & 30\arcdeg & No cone, $T=1500\,$K\\
NGC~5548 & 7 & $0.9-1.0$ & $10\arcdeg-30\arcdeg$ & Cone, $T=900\,$K and No cone, 
$T=1500\,$K\\
NGC~7172 & 6 & $0.7-1.6$  & $40\arcdeg$ & Cone, $T=900\,$K and 
No cone, $T=1500\,$K\\
NGC~7469 & 9 & 1.1 & $0\arcdeg-30\arcdeg$ & Cone, $T=900\,$K and No cone, 
$T=1500\,$K\\ 
Mkn~1    & 6 & $1.2$  & $40\arcdeg-64\arcdeg$ & Cone and No cone, 
$T=900\,$K\\
Mkn~348  & 7 & $0.4-1.0$  & $36\arcdeg-44\arcdeg$ & All\\
Mkn~533  & 7 & $1.5$ & $30\arcdeg-36\arcdeg$ & Cone and No cone, 
$T=900\,$K\\
Mkn~573  & 6 & $0.4-0.8$ & $36\arcdeg-50\arcdeg$ &  Cone and No cone, 
$T=900\,$K\\
MCG$-$5-23-16 & 5 & $0.7-0.8$ & $30\arcdeg-36\arcdeg$ & Cone and No cone, 
$T=900\,$K\\
IC~4329A & 6 & 1.4 & $10\arcdeg$ & Cone, $T=900\,$K\\
\enddata
\tablecomments{$\chi^2/\nu$ is $\chi^2$ per degree of freedom, where 
$\nu = n -2$ and $n$ is number of points for the SED.}
\end{deluxetable}

\begin{figure*}
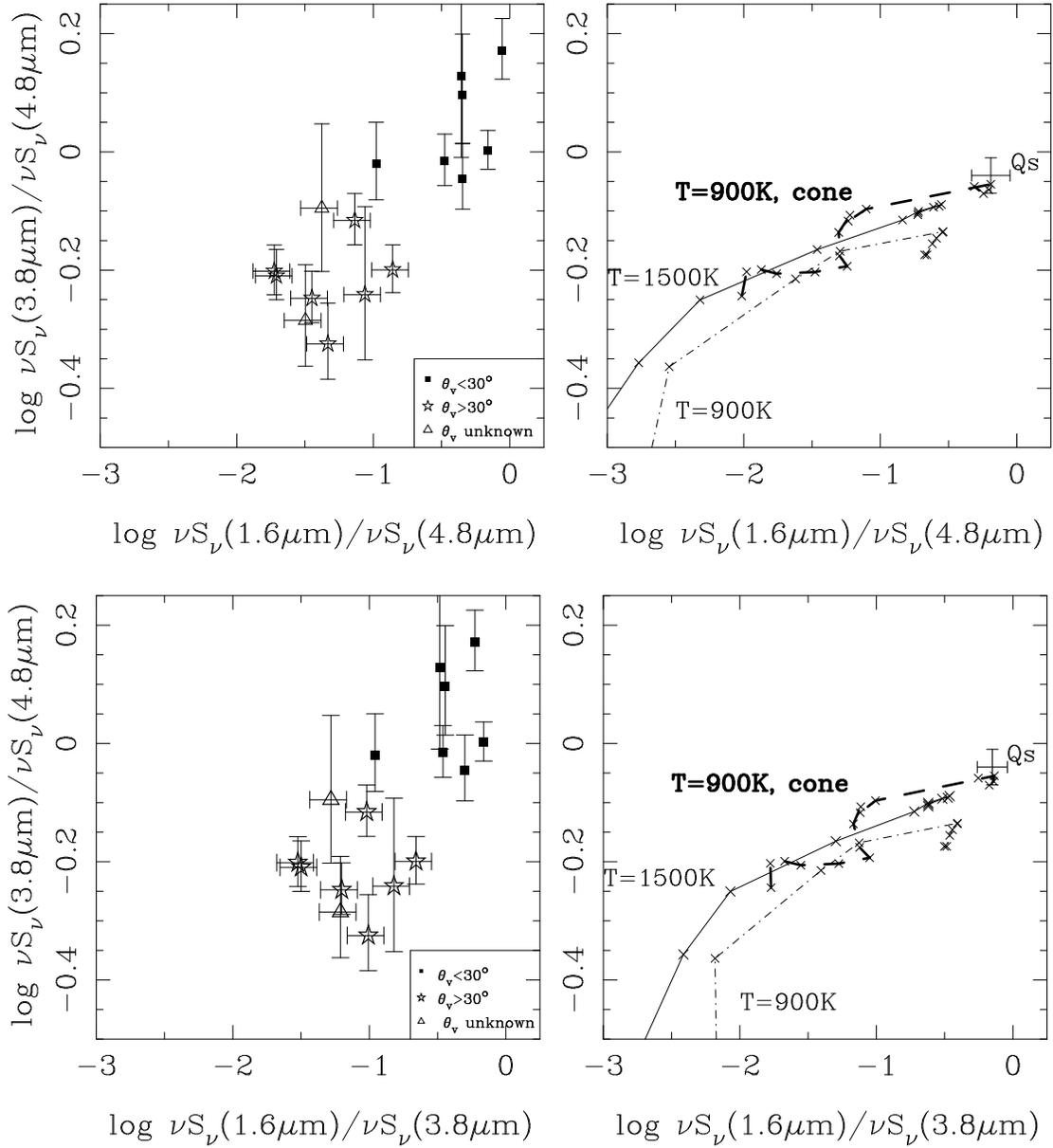

\figurenum{2}
\plotfiddle{alonso.fig2a.ps}{425pt}{-90}{70}{70}{-230}{660}
\plotfiddle{alonso.fig2b.ps}{425pt}{-90}{70}{70}{-230}{860}
\vspace{-15cm}
\caption{Non-stellar near-infrared color-color maps. The left panels
show the data for the sample of Seyfert galaxies, divided 
into galaxies with viewing angles $\theta_{\rm v} \leq 30\arcdeg$
and   $\theta_{\rm v} \geq 30\arcdeg$. 
The right panels show the outputs from Efstathiou et 
al. (1995) models for the torus+cone component model (thick
lines), and for the torus alone (coneless) model with two dust 
sublimation temperatures, $T=900\,$K (dot-dash lines) and
$T=1500\,$K (solid lines). The crosses on the lines indicate the viewing 
angles (see Table~6) which increase to the left. We also show
the average position of low redshift PG quasars (see text).}
\end{figure*}

\subsection{The Efstathiou \& Rowan-Robinson (1995) models}

For our analysis we consider the tapered 
disk model of Efstathiou \& Rowan-Robinson (1995). 
In this geometry, the height of the disk
increases with the radial distance, but tapers off to a constant
height in the outer part. We will use
a simple torus model (coneless) and a composite model with a torus and 
an optically thin cone. The latter model gave a better fit
to the near-infrared emission in NGC~1068 (Efstathiou et al. 
1995). The reasons for adding the 
optically-thin component rather than using 
a simple torus model to fit the SED of NGC~1068 were twofold. 
First, this additional 
optically thin cone component increases the continuum around
$3-5\,\mu$m, a condition necessary to fit the almost flat 
(in $\nu\,S_\nu$) observed SEDs of Seyfert 1s (see 
also below). This problem seems to affect all the 
current torus models (e.g., Pier \& Krolik 1993; Granato \& Danese 1994, 
and the model used here). Second, the simple torus model predicts 
strong absorption features at $9.7\,\mu$m for Seyfert 2s, which 
are not usually observed (Roche et al. 1991). See 
also the discussion in Efstathiou et al. (1995). 

The model 
assumes a maximum dust temperature (dust sublimation temperature) 
of $T=900\,$K, and the following 
parameters: $r_1/r_2=0.01$ (ratio of the inner and outer radii), 
$h/r_2=0.1$ (ratio of the height scale and the outer radius) and 
equatorial UV optical depth of $\tau_{\rm UV} = 1200$, 
with a half-opening
angle of the toroidal cone of $\theta_c = 30\arcdeg$.  
We will also explore for comparison purposes another coneless model with a
dust sublimation temperature of  $T= 1500\,$K.

A beaming factor 
was introduced by Efstathiou et al. 
(1995) to basically to boost the near-infrared flux, 
as the inclusion of conical dust by itself did not provide 
sufficient emission in this spectral range. 
The beaming factor is defined as the ratio of the 
radiation intensity
emitted by the continuum source towards the optically thin 
cone with respect to that directed towards the torus,  

\begin{equation}
f_{\rm b} = \frac{S_\nu (\theta < \theta_c)}
{S_\nu (\theta \ge \theta_c)}
\end{equation}

\noindent where $\theta_c$ is
the semi-opening angle of the torus, and $\theta$ is the angle 
measured from the pole. The beaming factor for this model 
is $f_{\rm b}=6$.

The viewing angles with $\theta_{\rm v} =0\arcdeg$ polar view, and 
$\theta_{\rm v} = 90\arcdeg$ equatorial view 
(note that in Efstathiou \& Rowan-Robinson
1995 models the authors measure the angles from the equator), 
UV optical depth, and corresponding visual extinctions of the 
coneless and cone torus model outputs are given in Table~6.

\subsection{SED fitting}

The first step was to scale the outputs from the model to the observed
SEDs in $\nu\,S_\nu$. The SED fitting was performed by minimizing the $\chi^2$ 
function of the models and the observations for the 
varying viewing angles to the AGN and normalizations at different
wavelengths, taking into account the errors due to the 
determination of the non-stellar fluxes and the photometric 
uncertainties. Note that we have not attempted to fit NGC~2992 or 
NGC~4968 because there are only four points for their SEDs.
In Table~7 we list the best fitting models (at 
$1\sigma$ level confidence) for the
SEDs where $n$ is the number of points in the SED, 
$\chi^2/\nu$ is $\chi^2$ per degree of freedom, 
$\theta_{\rm v}$
is the best fitting viewing angle (or range of angles) and the corresponding 
models. 

In Figure~1  we present the observed SEDs ($\log \nu\,S_\nu$ in 
units of ${\rm erg}\,{\rm cm}^{-2}\,{\rm s}^{-1}$) for the sample
of Seyferts along with the best fitting model scaled to 
the observed SED.  The model outputs are plotted for the 
viewing angles given  in Table~6 ($\theta_{\rm v} = 0\arcdeg$
is the top line, and $\theta_{\rm v}$ increases to the bottom). 
The abrupt transition 
seen in the model output at viewing angle $\theta_{\rm v}
=30\arcdeg$  is due to the assumed beaming factor. 
Using a beaming factor varying smoothly with $\theta$ would 
avoid the abrupt transition between 
those viewing angles.

As can be seen from Table~7 and 
Figure~1, the cone model provides good fits to the observed SEDs 
except in the cases of some intermediate SEDs 
(the SEDs of NGC~3227 and NGC~5506). The reason why  
the cone model fails to 
reproduce these intermediate SEDs is because of the abrupt transition 
produced for the assumed form of the beaming factor. If the 
beaming factor is decreased, then the cone model produces
an acceptable fit to the SEDs of these galaxies 
statistically indistinguishable from the coneless model.
In some cases we find from our SED fitting that both the cone and 
coneless models  produce
indistinguishable fits at the $1\sigma$ confidence level.

Despite the 
fact that we do not find compelling evidence to favor either the coneless 
or the cone model alone, we find that the viewing angle to the central 
source is very well 
constrained by the modelling of the SEDs. We find that 
none of the galaxies (not even the hard X-ray Compton-thick galaxies) 
in our sample show viewing angles greater 
than 64\arcdeg, with most of the obscured Seyferts displaying values of 
between 30 and 40\arcdeg. This does not mean that there are no Seyfert 
galaxies with $\theta_{\rm v} > 65\arcdeg$.
First, the values for the fitted viewing angles will 
depend on the assumed geometry for the torus model 
(mainly, the equatorial optical
depth and the beaming factor). Second, the shapes of the SEDs for 
viewing angles
above  $65\arcdeg$ become almost parallel, with the added complication 
that the fitted viewing angle is very sensitive near-infrared non-stellar 
estimates. Finally, we stress that our sample is not complete or unbiased in 
the sense that we are most likely not including very obscured objects. 

All the Seyfert 1s (see also Granato et al. 1997) except NGC~3227 
show relatively flat SEDs implying viewing angles in the  
range $\theta_{\rm v} = 0\arcdeg-30\arcdeg$ -- almost a face-on view. The 
coneless torus model does not account for the bright $1-5\,\mu$m
emission present in NGC~4151 and IC~4329A. Other models also invoke
additional near-infrared components in order to explain
the SED of NGC~1068. Granato et al. (1997) assumed a model 
similar to the 'anisotropic spheres' of Efstathiou \& Rowan-Robinson
(1995), whereas Pier \& Krolik (1993) used a hot component
made up of clouds inside the inner radius of the torus. 

As expected, the Seyfert 2 galaxies  show SEDs with different shapes
well explained with  the varying viewing 
angle to the central source. For example, NGC~5252 and NGC~5506 
(both classified as Seyfert 1.9) 
appear to be seen at  a viewing angle similar to that NGC~3227 (classified
as a Seyfert 1.5). For the particular model assumed here 
it would be $\theta_{\rm v} = 30\arcdeg$. The similarity 
between the SEDs of NGC~5506, NGC~5252 and NGC~3227 is also supported by
the detection in NGC~5506 and NGC~5252 of a broad line component in the 
infrared hydrogen recombination line Pa$\beta$ (Rix et al. 1990; 
Blanco, Ward, \& Wright 1990; and Ruiz et al. 1994, respectively). Veilleux 
et al. (1997), however, argued against the presence of 
such broad component in NGC~5506. 
Both galaxies also show relatively low hydrogen column densities 
derived from hard X-ray observations ($N_{\rm H} = 4.3\,{\rm and}\, 
3.4 \times 10^{22}\,{\rm cm}^{-2}$ for 
NGC~5252 and NGC~5506 respectively Bassani et al. 1999; equivalent
to an optical extinction of $A_V \simeq 22\,{\rm and}\,17\,$mag for the 
standard gas-to-dust conversion factor). NGC~5506, NGC~5252 and 
NGC~3227 appear as 
good examples of why averaging properties of type 1 and type 2 
objects may result in a loss of information.

\begin{figure*}
\figurenum{3}
\plotfiddle{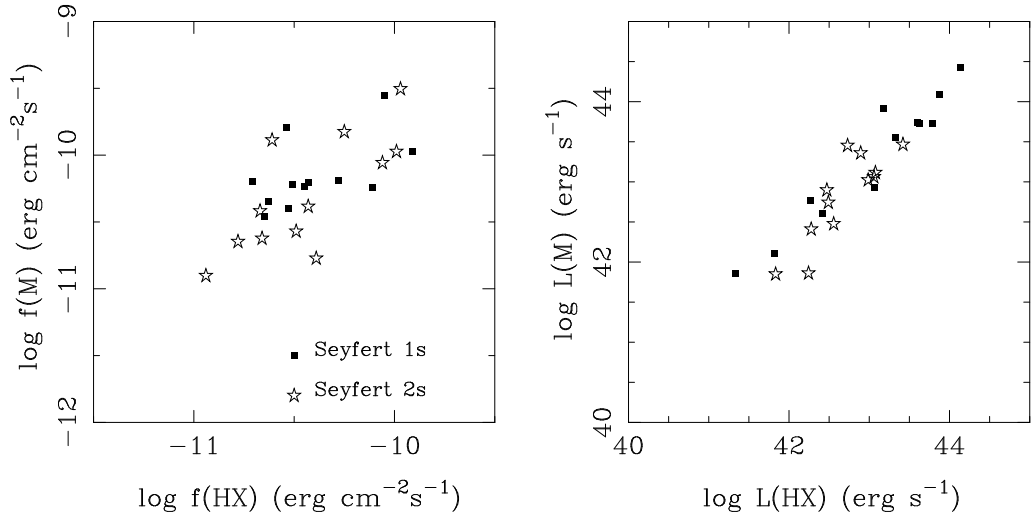}{425pt}{270}{70}{70}{-250}{630}
\vspace{-8cm}
\caption{$M$-band versus hard X-ray ($2-10\,$keV) correlations
in fluxes (left panel) and luminosities (right panel).}
\end{figure*}

\begin{figure*}
\figurenum{4}
\plotfiddle{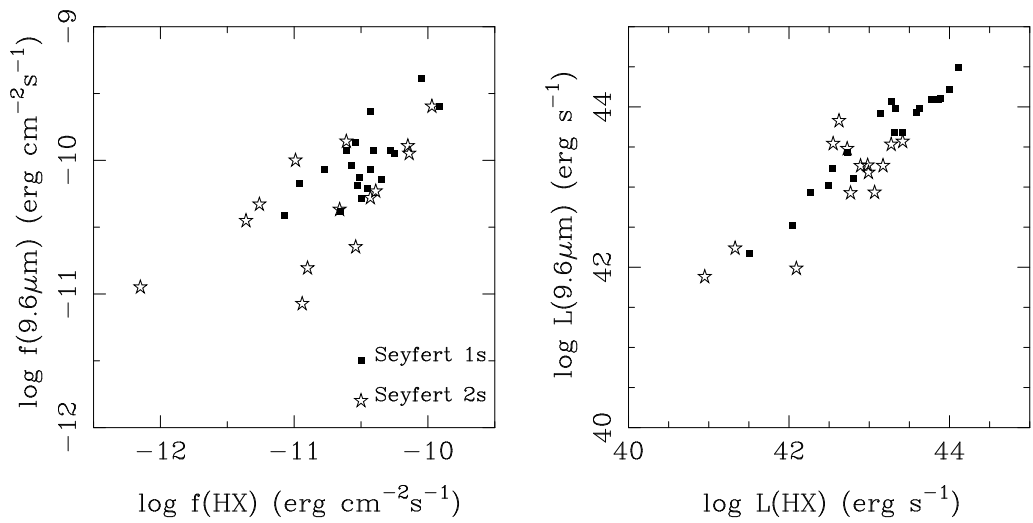}{425pt}{270}{70}{70}{-250}{670}
\vspace{-9cm}
\caption{As Figure~3, but with the ISO $9.63\,\mu$m fluxes and 
luminosities.}
\end{figure*}

The rest of the Seyfert 2s 
are viewed at angles $\theta_{\rm v} \geq 30\arcdeg$  
(NGC~1068, NGC~3281, NGC~7172, Mkn~348, Mkn~533, and
MCG$-$5-23-16). Of these galaxies Mkn~348, Mkn~533 and MCG$-$5-23-16 show
broad components in infrared hydrogen recombination lines
(Ruiz et al. 1994 for the first two 
and Veilleux et al. 1997 for MCG$-$5-23-16). Note
however, that Veilleux et al. (1997) disputed the finding 
of broad lines in Mkn~533. The optical
extinctions derived from hard X-ray observations 
(taken from Bassani et al. 1999)
vary from moderate (MCG$-$5-23-16, $A_V \simeq 8\,$mag),
intermediate (NGC~7172 $A_V \simeq 60\,$mag and NGC~3281 $A_V \simeq 400\,$mag)
to Compton-thick galaxies (NGC~1068 and Mkn~533). The most obscured 
SEDs in terms of the fitted viewing angles are Mkn~1 and  Mkn~573. 
Neither galaxy shows broad infrared hydrogen recombination lines
(Veilleux et al. 1997). 

We find a  tendency for the galaxies in our sample to have  
lower fitted  extinctions from the SED modelling than those derived from 
hard X-ray hydrogen column densities for those galaxies with 
relatively high values of $N_{\rm H}$. The particular values of the 
extinctions derived from fitting the SEDs depend on the assumed 
equatorial UV optical depth. As discussed in 
Efstathiou \& Rowan-Robinson (1995) in order to reproduce 
the high extinction needed for Compton thick objects, it would 
be necessary to increase the value of $\tau_{\rm UV}$ up to 4000 mag.
This would cause the models to produce a very deep  
$10\,\mu$m absorption feature. The main difficulty in tori 
with optical depths of the order of a few thousands, however, is that 
the SEDs become too narrow as the temperature drops more 
steeply with radius. This is evident in the models of Pier \&
Krolik (1993), and it is basically the reason why the 
'optically thin dust' in their model for NGC~1068
is cooler than ours. But even if we were to 
assume a much higher equatorial optical depth,   
the infrared SEDs of two Compton thick galaxies in our sample 
(NGC~1068 and Mkn~533) are well reproduced with 
relatively modest extinctions that could not possibly 
be as elevated as those inferred from hard X-ray column densities. 
The relatively low values of the extinction needed
for NGC~1068 were already noted in previous works (e.g.,
Efstathiou et al. 1995; Granato et al. 1997). We interpret the 
differing hard X-ray and infrared extinctions 
in the context of the torus + cone model. Most of 
the near-infrared emission seems to originate in the extended 
cone configuration which 
suffers less extinction than radiation emitted much closer
to the active galactic nucleus (e.g., the hard X-rays).

\subsection{Non-stellar color-color diagrams}
A number of works have attempted to use near-infrared to mid-infrared
colors or flux ratios to constrain the obscuring torus geometry, to 
derive viewing angles to the torus, or to simply quantify the different 
observed properties between type 1 and type 2 Seyferts. For instance,
Pier \& Krolik (1993) used the $L-N$ colors for Seyfert galaxies
to constrain the ratio $a/h$ (that is, the ratio of the inner radius
of the torus to its thickness, and therefore the opening
angle of the torus), but failed to reproduce the $L-N$
colors of Seyfert 2s, as well as the flat SEDs of Seyfert 1s and
PG quasars.  Murayama et al. (2000) have used Pier \&
Krolik (1993) models and the 
ratio of the $L$ ($3.5\,\mu$m) and IRAS $25\,\mu$m fluxes to 
derive torus properties such as the opening
angle and the critical viewing angle for which the broad 
line region (BLR) is visible. Fadda et al. 
(1998) compared the average of  number of near-infrared and mid-infrared 
colors for Seyfert 1s and 2s with predictions of a wide series of
models. These studies, as mentioned above, may suffer from unquantified 
large aperture effects in the mid-infrared where the underlying galaxy 
and/or star formation may make a significant contribution.

We explore non-stellar color-color diagrams involving observations 
with $\lambda < 5\,\mu$m for which we are certain about the non-stellar 
contribution. Due to the large apertures employed by the ISO 
instruments, there may be a significant stellar contribution. 
In Figure~2 we present non-stellar color-color 
diagrams: $L^\prime-M$ vs. $H-M$ 
[$\nu S_\nu(3.8\,\mu{\rm m})/\nu S_\nu(4.8\,\mu{\rm m})$ vs.
$\nu S_\nu(1.6\,\mu{\rm m})/\nu S_\nu(4.8\,\mu{\rm m})$, upper panel]
and $L^\prime-M$ vs. $H-L^\prime$ [$\nu S_\nu(3.8\,\mu{\rm m})/\nu 
S_\nu(4.8\,\mu{\rm m})$ vs.
$\nu S_\nu(1.6\,\mu{\rm m})/\nu S_\nu(3.8\,\mu{\rm m})$, bottom panel]. 
The left panels show  the sample of Seyfert galaxies. For the 
Seyfert 1s in Table~5
the $3.8\,\mu$m fluxes have been interpolated using the $3.5\,\mu$m
and $4.8\,\mu$m fluxes. The errors at $1.6\,\mu$m account for the uncertainty
associated with the non-stellar flux determination, whereas the errors
at $3.8\,\mu$m and $4.8\,\mu$m are due to the photometry 
and sky subtraction uncertainties. 
We have divided the sample of Seyferts  into 
galaxies with fitted viewing angles $\theta_{\rm v} \leq 30\arcdeg$
and $\theta_{\rm v} \geq 30\arcdeg$. NGC~2992 and NGC~4968 
are plotted with the symbol for unknown $\theta_{\rm v}$ since
they were not fitted in the preceding section.
In the right panels we show the outputs from the 
torus + cone model, and the coneless torus models with dust
sublimation temperatures $T=900, 1500\,$K. The crosses on the lines are 
the different viewing angles (increasing to the left). We also show 
the average position of nearby PG quasars using a near-mid infrared
spectral index of $\alpha = 1.4 \pm 0.13$
(Neugebauer et al. 1987). 

The effect of adding the cone component is readily seen from this figure. The 
cone model is able to reproduce the {\it bluer} 
(that is, flatter SEDs) $H-L^\prime$, 
$H-M$ and $L^\prime -M$  colors of the average PG quasar, 
and the Seyferts with $\theta_{\rm v}\leq 30\arcdeg$. Note that the 
three Seyfert 1s located in the 
upper right corner (NGC~3227, NGC~4151 and IC~4329A) have anomalous 
$M$-band fluxes (too low) as seen from the SEDs in Figure~1.
In addition, the coneless models produce much redder colors for large
values of the viewing angles, but these are not observed in 
our sample.

These two color-color diagrams provide a good means to separate 
Seyfert 2 galaxies with modest obscuration, $A_V$ up to $\simeq 
20\,$mag from hard X-ray observations (e.g., 
NGC~5252, NGC~5506 and MCG~$-$5-23-16, NGC~2992), 
from those with high obscuration. The two galaxies whose 
SEDs were not fitted in the preceding section also fit into 
these two categories. NGC~2992 displays a moderate 
absorption from hard X-ray observations, whereas 
NGC~4968 is a Compton-thick galaxy (Bassani et al. 1999). There appears 
to be a transition at around 
$\log \nu S_\nu(3.8\,\mu{\rm m})/\nu S_\nu(4.8\,\mu{\rm m}) \simeq
-0.15$. This may be interpreted as corresponding to a critical 
viewing angle for a relatively hard edge torus, above which 
the BLR becomes visible. This critical value is larger than 
the opening angle. The separation between the two classes is not so 
well defined if the 
$\log \nu S_\nu(1.6\,\mu{\rm m})/\nu S_\nu(4.8\,\mu{\rm m})$
(or $\log \nu S_\nu(1.6\,\mu{\rm m})/\nu S_\nu(3.8\,\mu{\rm m})$)
flux ratio is used instead, probably due to other effects such 
as foreground extinction still affecting the $1.6\,\mu$m 
fluxes.

\section{Mid-infrared emission of AGN, an indicator of the AGN
luminosity?}
\subsection{The mid-infrared versus hard X-ray correlations}
As mentioned in the 
introduction all torus models predict strong 
emission at mid-infrared wavelengths. 
Based on ISO observations of quasars it has been 
found that there is a good correlation between the 
infrared ($3-40\,\mu$m) luminosities and other properties 
(blue luminosity, soft X-ray luminosities) and this is 
indicative that an important fraction of the the mid-infrared energy 
is produced 
by dust heated by the AGN (Polletta et al. 2000; Haas et al. 
2000).  To test if this is the case for both Seyfert 1s and Seyfert 2s, it is 
necessary to have a good indicator of the AGN luminosity to which 
we can ratio the mid-infrared emission.

The hard X-ray ($2-10\,$keV) emission 
of Seyfert galaxies is known to be a good indicator of 
the intrinsic luminosity of the AGN for those cases where 
it is  transmitted through the torus, that is, in 
Compton thin galaxies. Other proposed indicators of the
AGN power of both Seyfert 1s and Seyfert 2s 
include the [O\,{\sc iii}]$\lambda$5007 luminosity 
(Mulchaey et al. 1994; Heckman 1995) or the non-thermal 
1.45\,GHz radio continuum
(Heckman 1995). The [O\,{\sc iii}] luminosities
may be affected by extinction, distribution of the gas in 
the parent galaxy, and star formation.   Heckman (1995) found that the  
$10.6\,\mu$m emission of Seyfert 2s when ratioed to the non-thermal
1.4\,GHz emission is four times smaller than that of Seyfert 1s, 
and this would suggest that even the mid-infrared emission of 
Seyfert 2s is anisotropic. However, caution is needed when interpreting 
Heckman's results; the radio emission of Seyfert 
galaxies may not be an isotropic property 
because of Doppler boosting of the core emission in Seyfert 1s.
Due to the uncertainties associated with the use of the 
radio and [O\,{\sc iii}] luminosities as indicators of the AGN power, 
we will test the above assumption using hard X-ray measurements.

We use the $M$-band ($4.8\,\mu$m) measurements presented in this 
work for Seyfert 2 galaxies, and for other Seyferts (mainly 
Seyfert 1s compiled by Ward et al. 1987), and the ISO $9.63\,\mu$m 
fluxes from Clavel et al. (2000), and hard X-ray ($2-10\,$keV) measurements 
(compiled by Mulchaey et al. 1994 and  Alonso-Herrero et al. 1997, and 
references therein) and 
new measurements from Bassani et al. (1999). The hard X-ray
fluxes of Seyfert 2s have been corrected for the intrinsic absorption.

In Figures~3 and 4 we plot the correlations 
in fluxes and luminosities (for $H_0 = 75\,{\rm km\, s}^{-1}
\,{\rm Mpc}^{-1}$) for the $4.8\,\mu$m and $9.63\,\mu$m 
measurements respectively. Both figures provide good evidence that the 
$5-10\,\mu$m emission in AGNs is powered by the central source. 
Note that in these two diagrams only Seyfert 2 galaxies  with relatively 
low attenuations for which the hard X-rays are transmitted 
(Compton thin, $N_{\rm H} \le 10^{24}\,{\rm cm}^{-2}$, 
or $A_V \lesssim 500\,$mag) are plotted. A similar correlation 
has been found by Krabbe, B\"oker, \& Maiolino (2000) using $N$-band 
(at $10.2\,\mu$m) imaging of a small sample of Seyfert galaxies.
These two correlations between the hard X-ray and the mid-infrared
luminosities are  very much improved even for Seyfert 2 galaxies 
when compared with shorter wavelengths ($2.2$ and $3.5\,\mu$m, 
Alonso-Herrero et al. 1997). 

We have performed a statistical
analysis of the correlations (both in fluxes and luminosities)
using the {\sc asurv} survival analysis program (Isobe \&
Feigelson 1990; LaValley, Isobe, \& Feigelson 1992). The 
results are tabulated in Table~8. For the $L$-band 
vs. hard X-ray correlations we have used the data for Seyfert 
1s and 2s presented in Alonso-Herrero et al. (1997). 
In this study we  demonstrated that for 
some Compton-thin Seyfert 2 galaxies much of the $L$-band 
($3.5\,\mu$m) emission is attenuated, and this fact was 
used to estimate the extinctions to the AGN. The reason 
for the improved correlations in the mid-infrared  ($4.8\,\mu$m and 
$9.6\,\mu$m) is the reduced extinction as the wavelength increases. 
As we saw in Section~4.2, in our sample there are no Seyfert galaxies 
with viewing angles $\theta_{\rm v} > 64\arcdeg$. This means that
we should not expect extinctions higher than $A_V=130\,$mag or
$A(4.8\,\mu{\rm m}) \simeq 3\,$mag (using Rieke \& Lebofsky
1985 extinction law), which is in good agreement with the observed
scatter in the $4.8\,\mu$m vs. hard X-ray correlation.

As discussed in Section~4.1, in the torus + cone model most of
the near-infrared emission is produced in the optically
thin cone. This component is located between the BLR
and the NLR. The conical dust is more visible 
from all viewing angles, and it is  probably illuminated 
by a stronger continuum (due to the beaming). This is confirmed by 
the findings of Bock et al. (1998) that 
NGC~1068 shows dust emission from the wall of the cavity, i.e., 
perpendicular to the axis of the torus, and in an extended
component ($\simeq 100\,$pc) coincident with the radio emission 
and ionization cone (Alloin et al. 2000). 
This naturally explains the 
good correlation between the 4.8 and $9.7\,\mu$m and the hard 
X-ray emissions. Moreover, this may offer an explanation for the 
Compton optically thick Seyferts ($N_H > 10^{24}\,$cm$^{-2}$)
that are bright in the mid-infrared. The examples in our
sample are NGC~1068, Mkn~533, and NGC~4968. 
This kind of objects are also present in the 
CfA sample (Quillen et al. 2001, in preparation). 
We conclude that the $5-10\,\mu$m
emission in Seyfert galaxies is a good indicator of the AGN power.

\begin{deluxetable}{lcccc}
\tablewidth{10cm}
\tablecaption{Correlation statistics for Seyfert 1s and Seyfert 2s.}
\tablehead{\colhead{Correlation}  & \colhead{quantity} &
\colhead{$N$} & \colhead{$r_{\rm s}$} & \colhead{probability}}
\startdata
$L$-band vs. HX & flux & 33 & 0.26 & $1.4 \times 10^{-1}$\\
$M$-band vs. HX & flux & 24 & 0.61 & $3.3 \times 10^{-3}$\\
$9.6\,\mu$m vs. HX & flux & 33 & 0.66 & $2.0 \times 10^{-4}$\\
\hline
$L$-band vs. HX & lum & 33 & 0.90 & $<7 \times 10^{-6}$\\
$M$-band vs. HX & lum & 24 & 0.94 & $6.8 \times 10^{-6}$\\
$9.6\,\mu$m vs. HX & lum & 33 & 0.90 & $6.7 \times 10^{-7}$\\
\enddata
\tablecomments{$N$ is the number of objects in the sample, $r_{\rm s}$ 
is the Spearman correlation rank, and prob is the probability that the
correlation is not present.}
\end{deluxetable}

\subsection{Mid-infrared counterparts of highly 
obscured X-ray galaxies}
The use of mid-infrared measurements 
as indicators of the AGN luminosity may have very important implications. 
Recent deep field surveys from Chandra have found
a number of hard X-ray sources which are not identified optically
(e.g., Mushotzky et al.~2000).
These sources are candidates for the highly absorbed objects
needed to account for the spectral index of the hard X-ray background
(Setti \& Woltjer 1989).  Some of the X-ray identified sources
require high levels of absorption to account for the flatness
of their X-ray spectrum, suggesting
that these sources might be opaque even at $10\,\mu$m.
A recent comparison of faint SCUBA sources with Chandra sources
found only one source in common (Fabian, Smail, \& 
Iwasawa 2000). However, in 
the preceding section we have shown that Compton-thick sources can 
be bright in the mid-infrared. If such sources are common then follow up 
deep surveys
can search for mid-infrared  counterparts to highly absorbed X-ray sources.

For the non-Compton thick sources ($N_{\rm H} \le 10^{24}\,{\rm cm}^{-2}$)
the $5\,\mu$m luminosity is similar to that at $2-10\,$keV (Figure~3).  
Mushotzsky et al. (2000)  found 
2000 and 300 sources per square degree with $2-10\,$keV fluxes greater
than $2 \times 10^{-15}$ and $10^{-14}$ erg cm$^{-2}$ s$^{-1}$ respectively.
These sources would correspond to fluxes of 3 and 17$\mu$Jy at rest wavelength
of $5\,\mu$m.  If the rest 10 and $20\,\mu$m luminosities are twice that of 
$5\,\mu$m flux then we would expect $13$ and $66\mu$Jy at a rest wavelength of
$10\,\mu$m and 26 and 130$\mu$Jy at a rest
wavelength of $20\,\mu$m for these sources.

The K correction in the mid-infrared is likely to make the observed sources
more difficult to detect since the spectrum is red out to $\sim 30\,\mu$m.
The obscured sources are most likely to be detected 
between $\sim 15-90\,\mu$m where the rest wavelength $5-30\,\mu$m
peak is redshifted into the observed wavelength range,
though unobscured sources would suffer less from the K correction.
Deep mid-infrared surveys planned with SIRTF will go deep enough
over large enough areas that 
mid-infrared candidates to absorbed X-ray sources should be found.
However we expect based on ISO surveys that in this wavelength
region the number of counts from starbursts will dominate by
at least an order of magnitude the counts from AGNs.
If these sources are at a redshift of 2.0
then the X-ray luminosity of a source with
hard X-ray fluxes $10^{-14}$ erg cm$^{-2}$ s$^{-1}$
is only $\sim 10^{44}$ erg s$^{-1}$
(assuming $q_0 =0.15$ and $H_0 = 75$) which implies that
these sources might just be Seyferts.
If they are L* galaxies they would be difficult to detect
in optical follow-up searches ($m_I > 26$).


\section{SUMMARY AND CONCLUSIONS}
We have presented $1-5\,\mu$m imaging for a sample of Seyfert 2 
galaxies. The data have been complemented with existing 
HST/NICMOS ($1-2.2\,\mu$m) and ISO ($6-15\,\mu$m) observations of 
the Seyfert 2s, as well as a small comparison sample of Seyfert 1s. 
We have performed a very careful estimate of the non-stellar fluxes 
in the $1-5\,\mu$m range to construct non-stellar infrared SEDs. We fit
the non-stellar infrared (up to 
$15\,\mu$m) SEDs with the Efstathiou \& Rowan-Robinson
(1995) code for tapered disks. We consider a simple torus model 
(coneless) and a composite model which includes
a torus and an optically thin cone component. The latter model was
found to fit well the SED of NGC~1068 (Efstathiou et al. 1995). 
From our analysis we find that:

\begin{itemize}

\item The non-stellar infrared SEDs for the Seyfert galaxies in our sample 
display a variety of shapes, which can all be well fitted by 
either cone or coneless models. Although the model fits to the
SEDs do not clearly favor the cone models, as claimed by 
Efstathiou et al. (1995), we nevertheless find that the
viewing angle towards the central source is well constrained 
even in cases where the fits to the cone and coneless 
models are equally valid. For all the galaxies in our sample
we find viewing angles $\theta_{\rm v} \leq 64\arcdeg$, with  most of 
the obscured galaxies showing $\theta_{\rm v} \simeq 40\arcdeg$.
The fitted viewing angles are however dependent on the assumed
torus geometry. 

\item The non-stellar $L^\prime-M$ vs 
$H-L^\prime$ and $H-M$ color-color diagrams provide a good
means to separate those Seyfert 2s with 
moderate obscurations ($A_V \lesssim 20\,$mag from 
hard X-ray observations) from those with high 
obscuration. The $L^\prime-M$,  
$H-L^\prime$ and $H-M$ colors of Seyfert 1s and Seyfert 2s with 
viewing angles of $\theta_{\rm v} < 30\arcdeg$ are better 
reproduced with the cone model.

\item The extinctions derived from the model fits to the SEDs tend
to be less than those inferred from measurements of the hard
X-ray attenuations. This can be understood if the material 
responsible for obscuration of the infrared continuum is in a
different location from the gas column causing absorption 
of the X-rays. 

\item There is a good correlation between $4.8\,\mu$m 
and ISO $9.7\,\mu$m and hard X-ray fluxes and luminosities 
for both Seyfert 1s and Compton thin ($N_{\rm H} 
\le 10^{24}\,{\rm cm}^{-2}$) Seyfert 2s. The improved correlations 
at $4.8$ and $9.7\,\mu$m  with respect to those at shorter wavelengths 
are explained in terms of the reduced extinction.
Some Compton thick sources (e.g., NGC~1068 and 
Mkn~533) are bright infrared sources suggesting
that the component responsible for the bulk of the infrared 
emission in Seyfert galaxies is more visible from all viewing  
angles than that responsible for the hard X-ray emission. We conclude that the
mid-infrared emission in Seyfert galaxies can be used 
as a measure of the AGN luminosity. This has important 
implications for future mid-infrared searches of visually obscured 
objects which may make an important contribution 
to the hard X-ray background.

\end{itemize}

\section*{Acknowledgments}

We are grateful to Dr. P. Martini for providing us with some of the 
NICMOS images
used in this work prior to their publication. We also thank Dr. M. 
Ruiz for useful discussions.
AA-H was partially supported by the National
Aeronautics and Space Administration on grant NAG 5-3042 through the
University of Arizona.  AA-H thanks the Department of Physics 
and Astronomy, University of Leicester for their warm hospitality.

\end{document}